\documentclass[12pt,preprint]{aastex}

\newcommand\msun{\hbox{\,M$_\odot$}}
\newcommand\rsun{\hbox{\,R$_\odot$}}
\newcommand\lsun{\hbox{\,L$_\odot$}}
\newcommand\kms{ km~s$^{-1}$}

\newcommand\alfoph{$\alpha$~Oph~}
\newcommand\alfcep{$\alpha$~Cep~}

\def\kms{\ifmmode{\rm km\thinspace s^{-1}}\else km\thinspace s$^{-1}$\fi}

\shorttitle{Rapid rotators: $\alpha$ Oph \& $\alpha$ Cep}
\shortauthors{Zhao et al.}

\begin{document}

\title{Imaging and Modeling Rapidly Rotating Stars: $\alpha$~Cephei and $\alpha$~Ophiuchi }

\author{M.~Zhao\altaffilmark{1},
J.~D.~Monnier\altaffilmark{1},
E. Pedretti\altaffilmark{2},
N. Thureau\altaffilmark{2},
A. M{\'e}rand\altaffilmark{3,4},
T. ten Brummelaar\altaffilmark{3},
H. McAlister\altaffilmark{3},
S. T. Ridgway\altaffilmark{5},
N. Turner\altaffilmark{3},
J. Sturmann\altaffilmark{3},
L. Sturmann\altaffilmark{3},
P. J. Goldfinger\altaffilmark{3},
C. Farrington\altaffilmark{3}
}

\altaffiltext{1}{mingzhao@umich.edu: University of Michigan, Astronomy Department,
941 Dennison Bldg, Ann Arbor, MI 48109-1090, USA}
\altaffiltext{2}{University of St. Andrews, Scotland, UK}
\altaffiltext{3}{The CHARA Array, Georgia State University}
\altaffiltext{4}{European Southern Observatory}
\altaffiltext{5}{National Optical Astronomy Observatory, NOAO, Tucson, AZ}

\begin{abstract}
We present sub-milliarcseond resolution imaging and modeling of two nearby rapid rotators $\alpha$ Cephei and $\alpha$ Ophiuchi, obtained with the CHARA array - the largest optical/IR interferometer in the world. Incorporating a gravity darkening model, we are able to determine the inclination, the polar and equatorial radius and temperature, as well as the fractional rotation speed of the two stars with unprecedented precision. 
The polar and equatorial regions of the two stars have $\sim$2000K temperature gradient, causing their apparent temperatures and luminosities to be dependent on their viewing angles.
Our modeling allow us to determine the true effective temperatures and luminosities of $\alpha$ Cep and $\alpha$ Oph, permitting us to investigate their true locations on the H-R diagram.  
These properties in turn give us estimates of the masses and ages of the two stars within 
a few percent of error using stellar evolution models.  Also, based on our gravity darkening modeling, we propose a new method to estimate the masses of single stars in a more direct way through V$\sin i$ measurements and precise geometrical constraint.  
Lastly, we investigate the degeneracy between the inclination and the gravity darkening coefficient, which especially affects the modeling of $\alpha$ Oph. Although incorporating V$\sin i$ has lifted the degeneracy to some extent, higher resolution observations are still needed to further constrain the parameters independently.

\end{abstract}


\section{Introduction}
\label{introduction}
In the past few years, optical interferometers have resolved the elongated photospheres of rapidly-rotating stars for the first time. The emergence of these high angular resolution observations of hot stars has shined a spotlight on critical areas of stellar evolution and basic astrophysics that demand our attention. For decades, stellar rotation was generally overlooked in stellar models and was regarded to have a trivial influence on stellar evolution because most stars are slow rotators, such as the Sun \citep{Maeder2000}.  Although the 
effects of rotation on solar type stars are indeed relatively mild, they are more prominent on hot stars.  Studies have shown that a large fraction of hot stars are rapid rotators with rotational velocities more than 120 \kms \citep{Abt1995, Abt2002}. Virtually all the emission-line B (Be) stars are rapid rotators 
with rotational velocities of $\sim90\%$ of breakup \citep{Fremat2005}.
Stars that are rapidly rotating have many unique characteristics. The centrifugal force from rapid rotation distorts their 
photospheres and causes them to be oblate. This distortion causes their surface brightness and $T_{eff}
$ to vary with latitude, and their equatorial temperatures are predicted to be much cooler than their polar temperatures, a 
phenomenon known as ``Gravity Darkening" \citep{vonZeipel1924, vonZeipel1924b}. Recent stellar models that took rotation into account showed that rapid rotation also affects stars' 
luminosity, abundance \citep{Pinsonneault1997}, evolution, and increases their lifetime \citep{Kiziloglu1996, Talon1997, Meynet2000}. It is also linked to stellar wind, mass loss \citep[e.g., ][]{Maeder2007}, and
 even Gamma-Ray bursts \citep{MacFadyen1999, MacFadyen2001, Burrows2007}.

The development of long baseline optical interferometry in recent years has evoked observations on 
several nearby rapid rotators, for instance, Altair, Vega, Achernar, Alderamin ($\alpha$ Cephei) and Regulus \citep{vanBelle2001,Aufdenberg2006, Peterson2006b, Domiciano-de-Souza2003, vanBelle2006, McAlister2005, Kervella2006, Monnier2007sci}. These studies confirmed the general picture of 
von Zeipel's gravity darkening law, 
but also raised discrepancies between observations and the widely adopted standard von Zeipel model (i.e., $T_{eff}\propto g_{eff}^{\beta}$, where $\beta$ is the gravity darkening coefficient, and $\beta= 0.25$ for fully radiative envelopes).  Particularly, the recent study of \citet{Monnier2007sci} on Altair
showed that their model prefers a non-standard gravity darkening law. What is more interesting is that they 
reconstructed a model-independent image for Altair and found a darker-than-expected equator 
 compared to the model. This suggests for the first time from observations that the standard gravity darkening law may work
  only at a basic level and other mechanisms need to be introduced to account for the extra darkening. To 
  address this issue, we will need more detailed studies and model-independent images of rapid rotators.

In this paper, we present our study of the two nearby rapid rotator $\alpha$ 
Cephei and $\alpha$ Ophiuchi, observed with the CHARA long baseline optical/IR interferometer array and the MIRC beam combiner.  
The star $\alpha$ Cephei (\alfcep, Alderamin, HR 8162, $V$=2.46, $H$=2.13, d=14.96pc) is the eighth nearest A star in the sky. It was classified as an A7 IV-V star in early studies, but was recently classified as an A8V main sequence star by \citet{Gray2003}.  It is one of the few A stars (including Altair) that are found to have chromosphere activities \citep{Walter1995, Simon1997, Simon2002}. The V$\sin i$ measurements of \alfcep show large scatter, spanning from $\sim180 \kms$ to $\sim245 \kms$  \citep{Bernacca1970, Uesugi1970, ROYER2007,  Abt1995}. Recently, \citet{vanBelle2006} studied \alfcep using the CHARA array and found it 
 is rotating close to break-up, and its photosphere is elongated due to rapid rotation. 
 
 The star $\alpha$ 
Ophiuchi (\alfoph, Rasalhague, HR 6556, $V$ = 2.09, $H$=1.66, d=14.68pc) is a nearby subgiant binary system \citep{Wagman1946, Lippincott1966}, and is the seventh nearest A star in the sky. The primary is a A5IV sub-giant which was first identified as a class III star but was later corrected
  to class IV by \citet{Augensen1992} and \citet{Gray2001}. Several groups have tried to study the orbit of the system 
  \citep[][etc.]{McAlister1984, Kamper1989, Mason1999, Augensen1992, Gatewood2005}, and it was lately determined to have a period of $\sim8.6$ yrs and a semi-major axis between 0.4'' - 0.5''.  
  The mass determination of the primary has large scatter, ranging from 2\msun~ to 4.9\msun~ \citep[e.g,][]{Kamper1989, Augensen1992, Gatewood2005}. The companion, which is approximately a K2V star, is thought to have a mass of 
  0.5-1.2\msun \citep{Kamper1989, Augensen1992, Gatewood2005}, and is observed to be 3.5 mag fainter than the primary in the $K$ band \citep{Boccaletti2001}. 
  The size of the primary was estimated to be $\sim1.6$ - $1.7$\rsun 
  \citep{Barnes1978, blackwell1980}. Its rotational velocity V$\sin i$  ranges from 210\kms to 240\kms \citep{Bernacca1970, Uesugi1970, 
   Abt1995, Royer2002}, implying \alfoph is spinning at a significant fraction of its break-up speed of $\sim270 \kms$.

This paper is organized as follows. We report our observations and data reduction schemes in \S\ref{obs}.  We discuss our aperture synthesis imaging for \alfcep and \alfoph in \S\ref{imaging} and  present gravity darkening models for both of them in \S\ref{model}. In \S\ref{Teff}, we present their temperatures, luminosities, and their locations on the H-R diagram. Based on our modeling, we propose a new method to estimate the mass of a star in \S\ref{new_method}. Finally, we discuss our results in \S\ref{discuss} and present our conclusions in \S\ref{conclusion}.

\section{Observations and data reduction}
\label{obs}
Our observations were conducted at the Georgia State University (GSU) Center
 for High Angular Resolution Astronomy (CHARA) interferometer array along with the MIRC combiner. The
 CHARA array, located on Mt. Wilson and consisting of six 1-meter telescopes, is the longest optical/IR 
 interferometer array in the world \citep{Brummelaar2005}. The array is arranged in a Y-shaped 
 configuration and has 15 baselines ranging from 34m to 331m, providing resolutions up to $\sim$0.5 mas at the $H$ band and $\sim0.7$mas at the $K$ band. 
 
The Michigan Infra-Red Combiner (MIRC) was used here to combine  4 CHARA telescopes together for true interferometric imaging in the $H$ band, providing 6 visibilities, 4 closure phases and 4 triple amplitudes simultaneously in 8~narrow spectral channels \citep[see][for details]{Monnier2004,Monnier2006}. MIRC is designed for stable calibrations and precise closure phase measurements. It uses single mode fibers to spatially filter the light coming from the CHARA beams. The fibers are brought together by a V-groove array in a non-redundant pattern. The outgoing fiber beams are then collimated by a lenslet array and are focused by a spherical mirror to form an interference pattern, which consists of six overlapping fringes with non-redundant spatial frequencies.  The fringes are focused again by a cylindrical lens into a ``line" of fringes and are dispersed by low spectral resolution prisms with R $\sim50$. 
The dispersed fringes are finally detected by a PICNIC camera, where they fall onto 8 spectral channels spanning the $H$ band ($\lambda$=1.5 - 1.8 $\mu m$) \citep{Monnier2004, Monnier2006}. A detailed description of the control system and software can be found in \citet{Pedretti2009}.

The system visibilities of MIRC are very stable due to our use of single mode fibers. However,  the atmospheric turbulence changes faster than the 5.5ms readout speed of the camera, causing decoherence of the fringes that needs to be calibrated. We therefore observe several calibrators adjacent to our targets over each observing night. For the purpose of bias subtraction and flux calibration, each set of fringe data is bracketed with measurements of background (i.e., data taken with all beams closed),  shutter sequences (i.e., data taken with only one beam open at a time to estimate the amount of light coming from each beam), and foreground (i.e., data taken with all beams open but without fringes) \citep{Pedretti2009}. Each object is observed for multiple sets. During the period of taking fringe data, a group-delay fringe tracker is used to track  the fringes  \citep{Thrueau2006}. In order to track the flux coupled into each beam in ``real time" to improve the visibility measurements, we  use spinning choppers to temporally modulate the light going into each fiber simultaneously with fringe measurements. The chopper speeds were set to 25Hz, 30Hz, 35Hz and 40Hz in 2006 and were increased to 55Hz, 65Hz, 75Hz and 85Hz in 2007 to avoid overlap of modulating frequencies caused by chopper drifts. 

\begin{figure}[thb]
\begin{center}
{
\includegraphics[angle=0,width=3in]{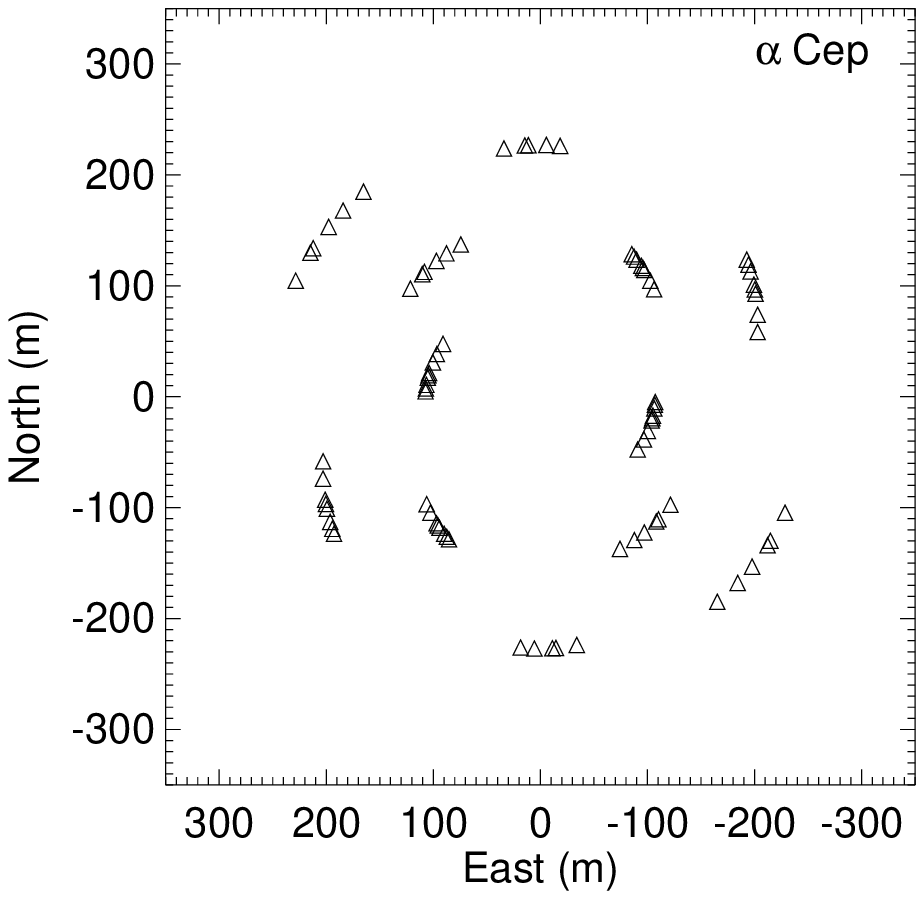}
\includegraphics[angle=0,width=3in]{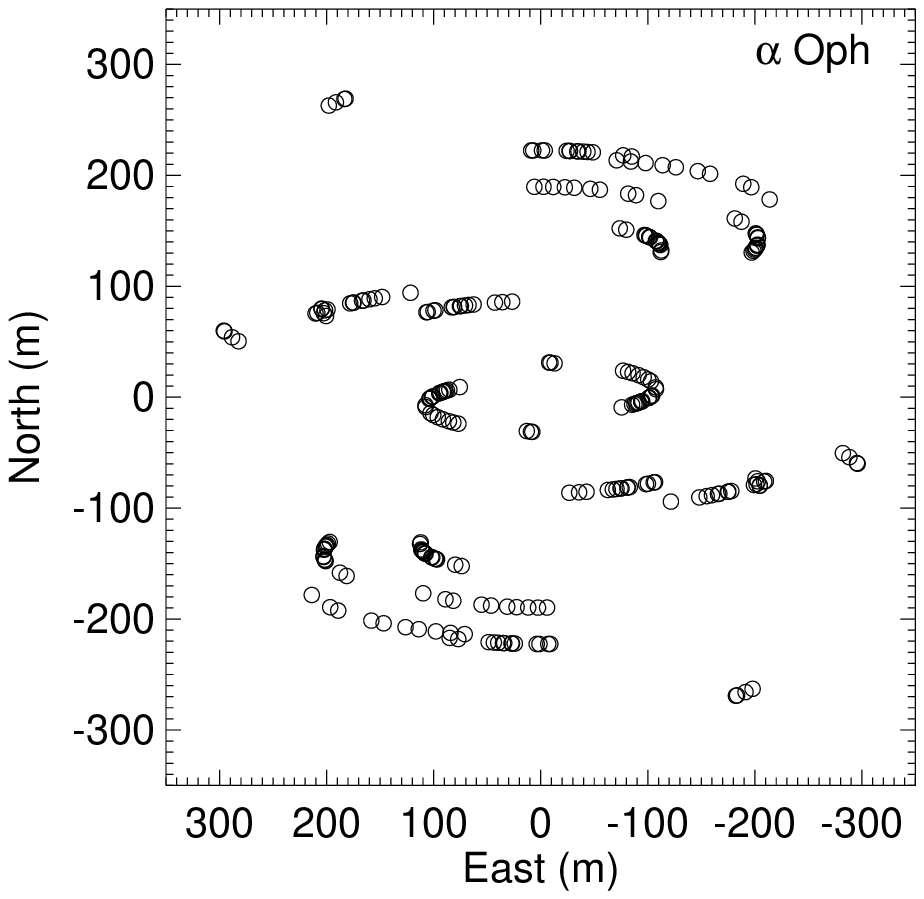}}
\hphantom{.....}
\caption{ 
Baseline coverages for \alfcep and $\alpha$ Oph. The longest baselines in the observations are 251m and 329m for \alfcep and $\alpha$ Oph, corresponding to resolutions of 0.68mas and 0.52mas respectively. The UV coverage can be obtained by dividing these two plots by corresponding wavelengths.  
\label{uv}}
\end{center}
\end{figure}

We  observed \alfcep on 4 nights in 2006 and observed \alfoph on 8 nights in 2006 and 2007, using various array configurations optimized for equal Fourier coverage in all directions for good imaging. The detailed log of our observations is listed in Table \ref{obslog}. Figure \ref{uv} shows the overall baseline coverage of our observations of \alfcep and $\alpha$ Oph. 

The data reduction process follows the pipeline outlined by \citet{Monnier2007sci}, which was validated using data on the binary 
$\iota$ Peg. In brief, after frame-coadding, background subtraction and Fourier transformation of the raw data, fringe amplitudes and 
phases are used to form squared-visibilities and triple products. Raw squared-visibilities are then estimated from the power 
spectrum after foreground bias subtraction. After the fiber coupling 
efficiencies are estimated using either the chopping signal or direct fit to the fiber profiles,  we obtain 
uncalibrated squared-visibilities and complex triple amplitudes. Finally, calibrators with known sizes are used to calibrate the drifts in overall system response before we obtain the calibrated squared-visibilites,
 closure phases, and complex triple amplitudes. The adopted sizes of our calibrators are listed in 
Table \ref{cals}. Corresponding errors of  the data are estimated by combining both the scatter of the data and calibration errors. 

\section{Aperture Synthesis Imaging}
\label{imaging}

We employed the publicly-available application ``Markov-Chain Imager for Optical Interferometry (MACIM)" \citep{Ireland2006} to reconstruct images for \alfcep and $\alpha$ Oph. The application applies the Maximum Entropy Method (MEM) \citep{Narayan1986} widely used in radio synthesis imaging, and has been validated on other test data \citep{Lawson2006}. Since the photosphere of a star has a sharp emission cut-off at the edge, which is imprinted in the highest spatial frequencies that cannot be observed, we constrain the field of view of the images within an ellipse to avoid spreading-out of the flux by the MEM procedure at the edge of the star. This constraint is appropriate for \alfcep and \alfoph due to their lack of any circumstellar emission outside of their photospheres. The details of this approach can be found in \citet{Monnier2007sci}. The ellipse prior is found by conducting MACIM imaging on a grid of $\sim400$ different ellipses with uniform surface brightness, spanning a range of possible sizes, axial ratios, and position angles. 
To ensure the smoothness of the image, 
we also de-weighted the high resolution data with
a gaussian beam of 0.3 milliarcsec FHWM, an approach usually applied in radio synthesis imaging.
The image with the global maximum entropy is then taken as the final result. 
We treated each wavelength channel as providing a distinct set of (u, v) plane coverage, ignoring any wavelength-dependence of the image itself. This assumption is well justified for \alfcep and \alfoph since the brightness profiles of their photospheres are almost identical in all channels in the $H$ band.



\begin{figure}[thb]
\begin{center}
{
\includegraphics[angle=90,width=3.2in]{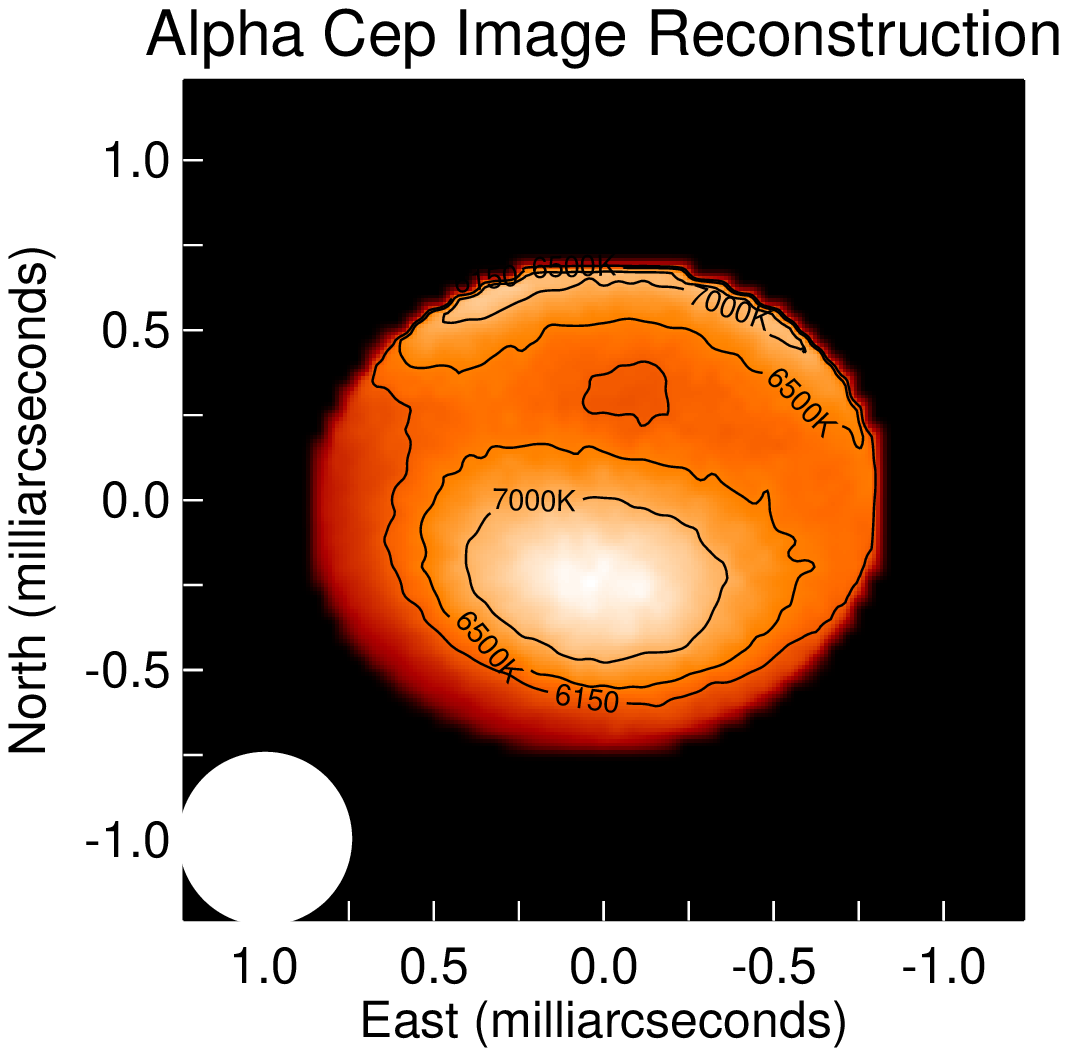}
\includegraphics[angle=90,width=3.2in]{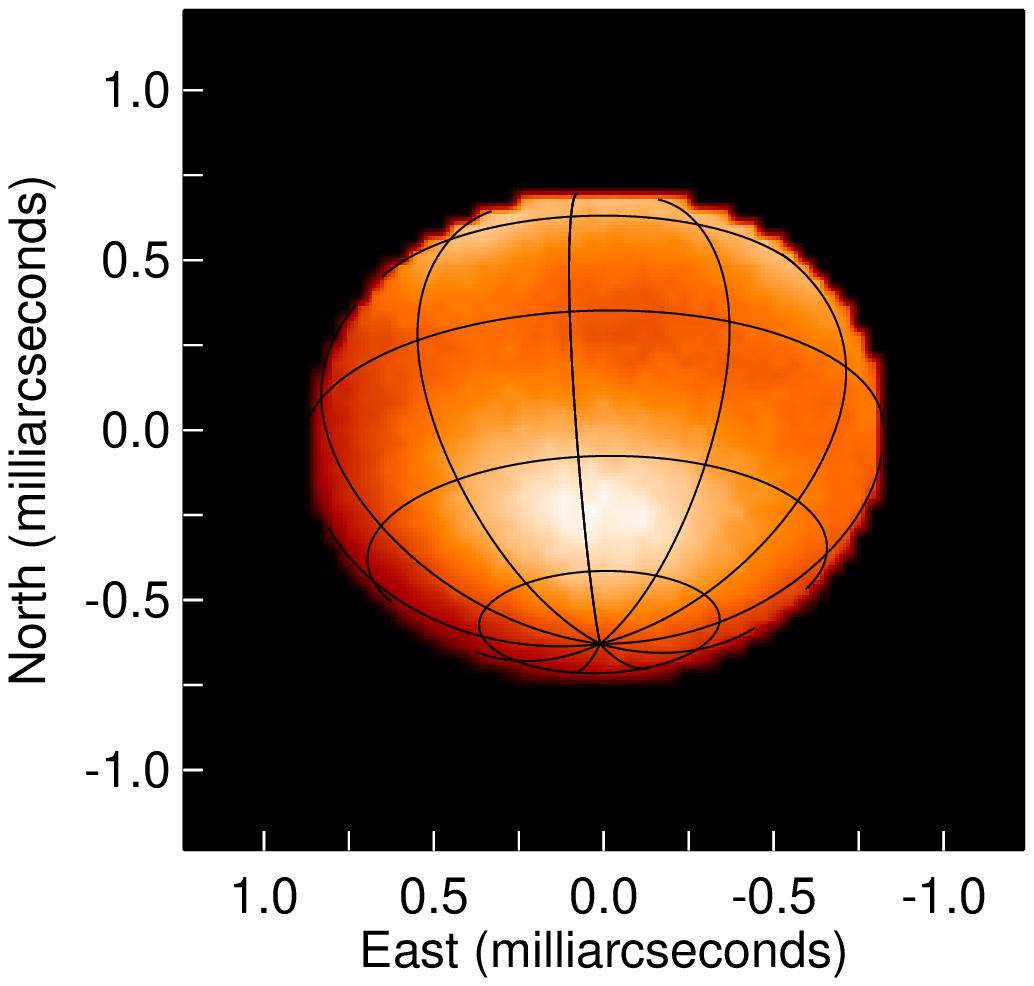}}
\hphantom{.....}
\caption{ 
Reconstructed MACIM image of $\alpha$ Cep. The left panel shows the contours of local brightness temperature. To help visualize the geometry of $\alpha$ Cep, the right panel shows its latitude and longitude using the positions from the standard model discussed in \S\ref{model}. The white circle at the bottom-left corner of the left panel shows the size of the convolving beam that we use for the image reconstruction. The total $ \chi^2_{\nu} $ of the image is 1.10. The resolution of the image is 0.68 milliarcsec.
\label{alfcep_img}}
\end{center}
\end{figure}
\begin{figure}[thb]
\begin{center}
{\includegraphics[angle=0,width=3.2in]{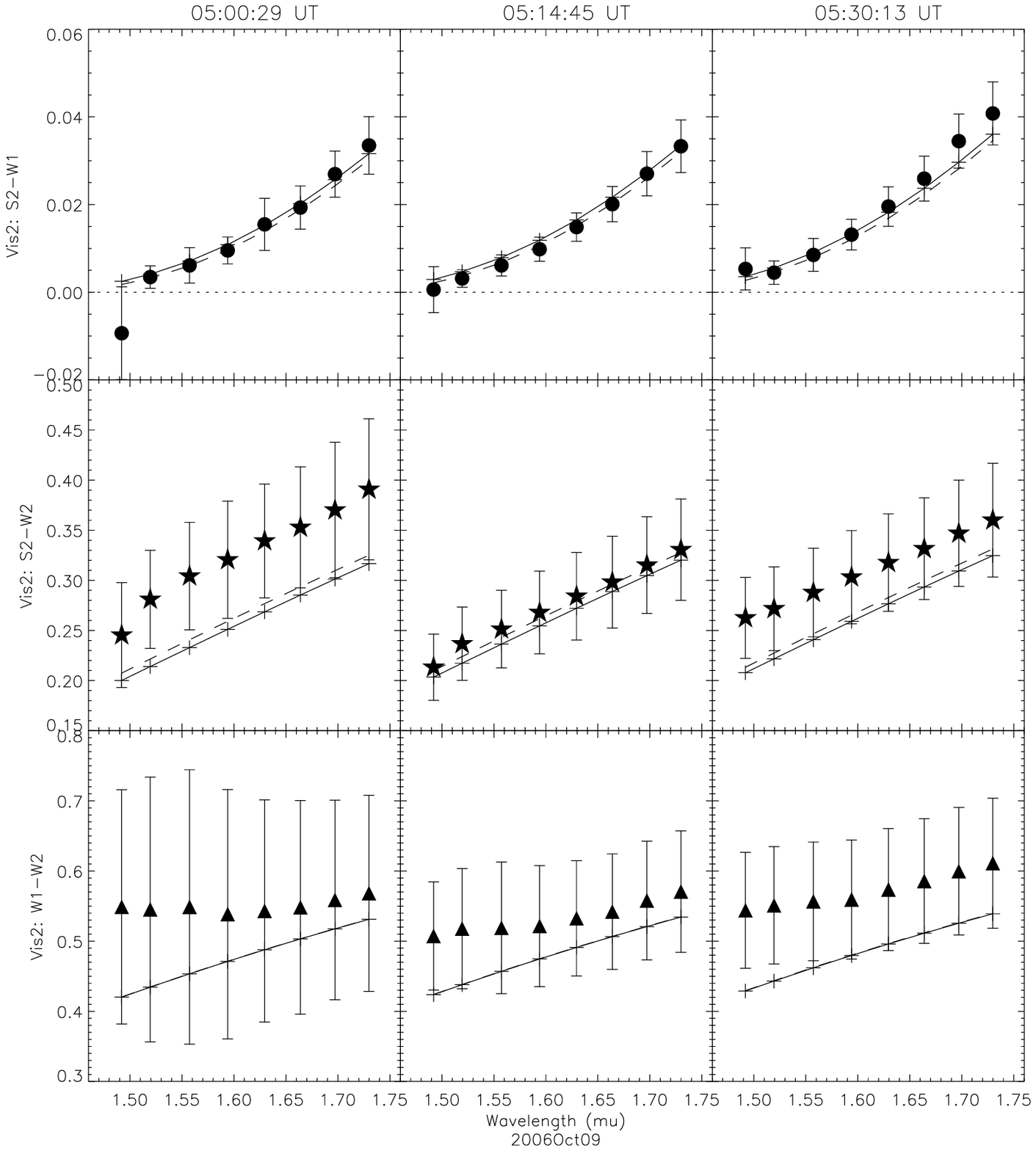}
\includegraphics[angle=0,width=3.2in]{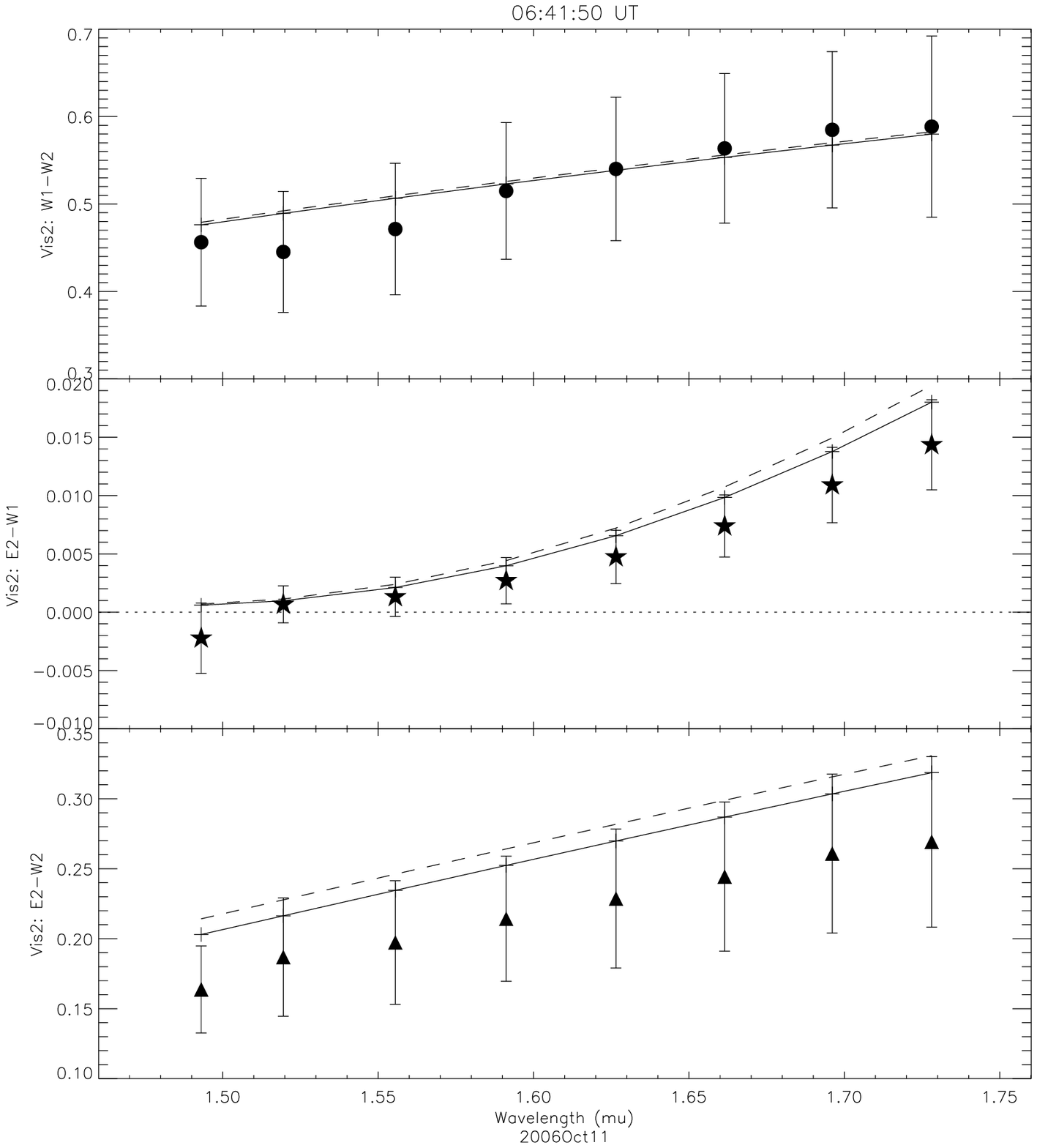}
\includegraphics[angle=0,width=3.2in]{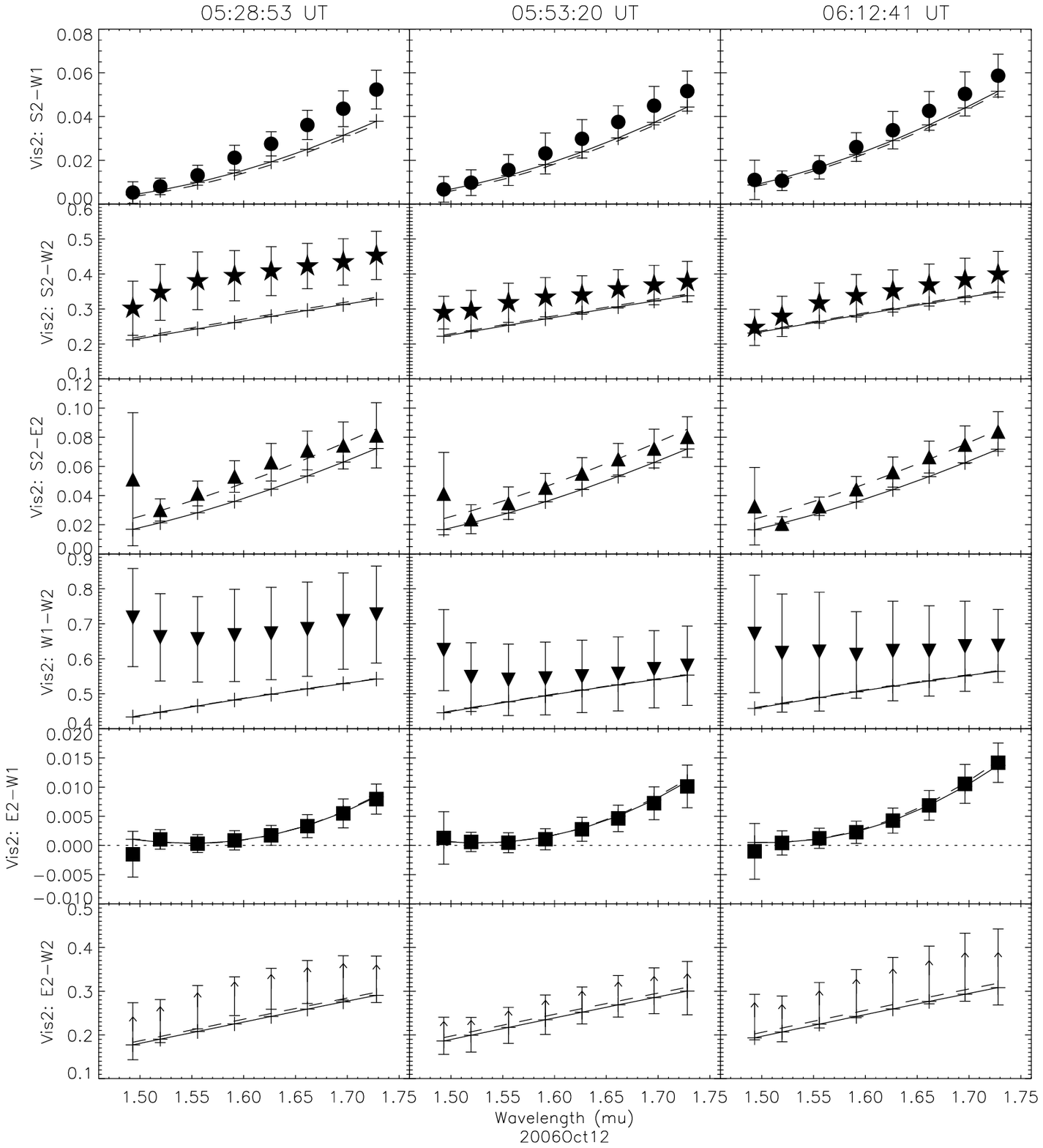}
\includegraphics[angle=0,width=3.2in]{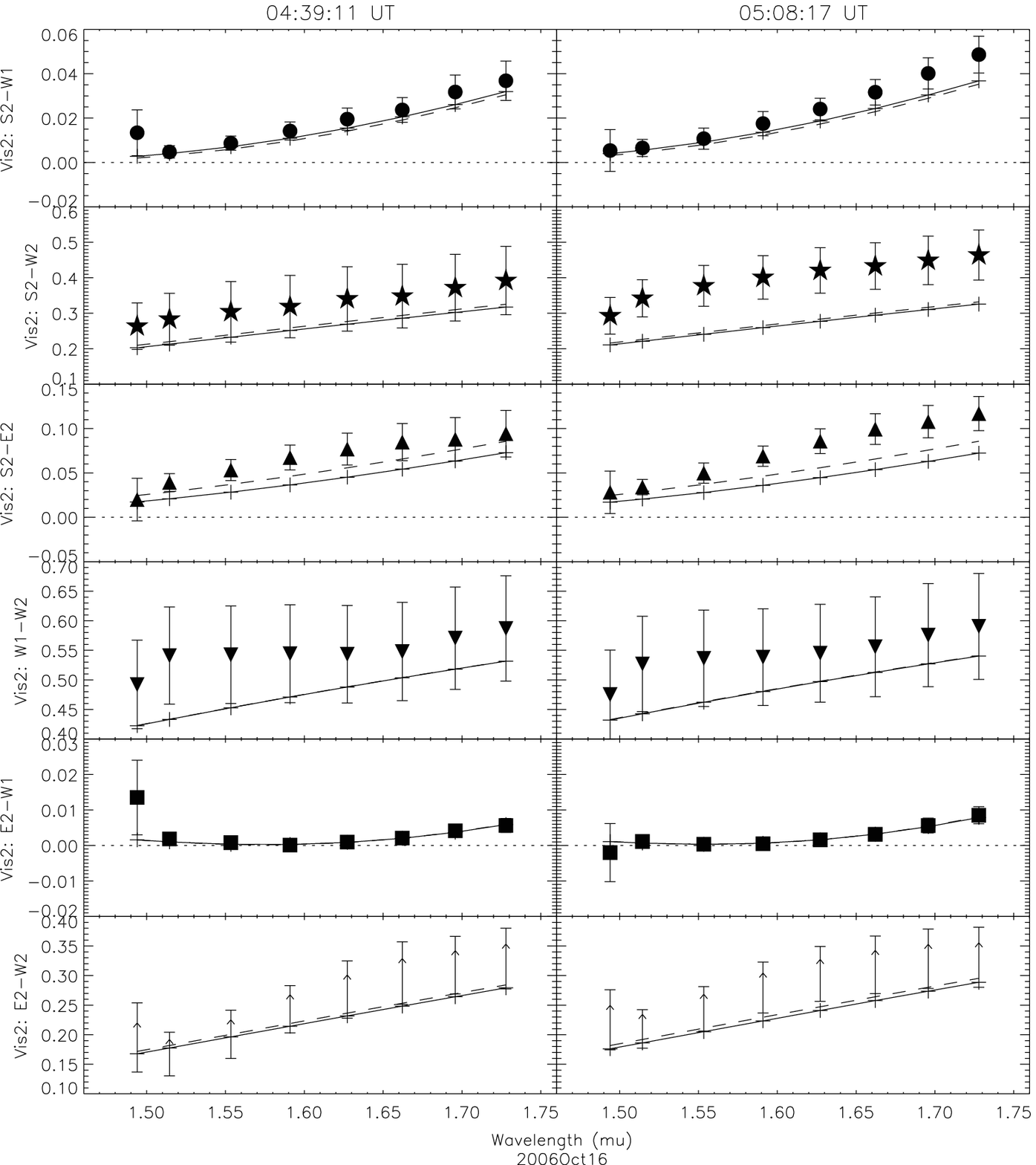}
}
\hphantom{.....}
\caption{ 
\alfcep squared-visibilities from the MACIM image (solid lines) and the gravity darkening model ($\beta=0.216$, dashed lines) vs. data (filled points with error bars). All four nights (2006 Oct09, 11, 12, 16) are shown here.
The $ \chi^2 _{\nu} $ of the image's squared-visibilities is 0.87, while that of the model is 0.80.
Each row stands for a different baseline, while the columns indicate different times of observation. The eight data points in each panel indicate the eight spectral channels 
of MIRC across the $H$ band. (Please refer to the electronic edition if the type size is too small.)
\label{alfcep_vis2_img}}
\end{center}
\end{figure}
\begin{figure}[thb]
\begin{center}
{\includegraphics[angle=0,width=3.2in ]{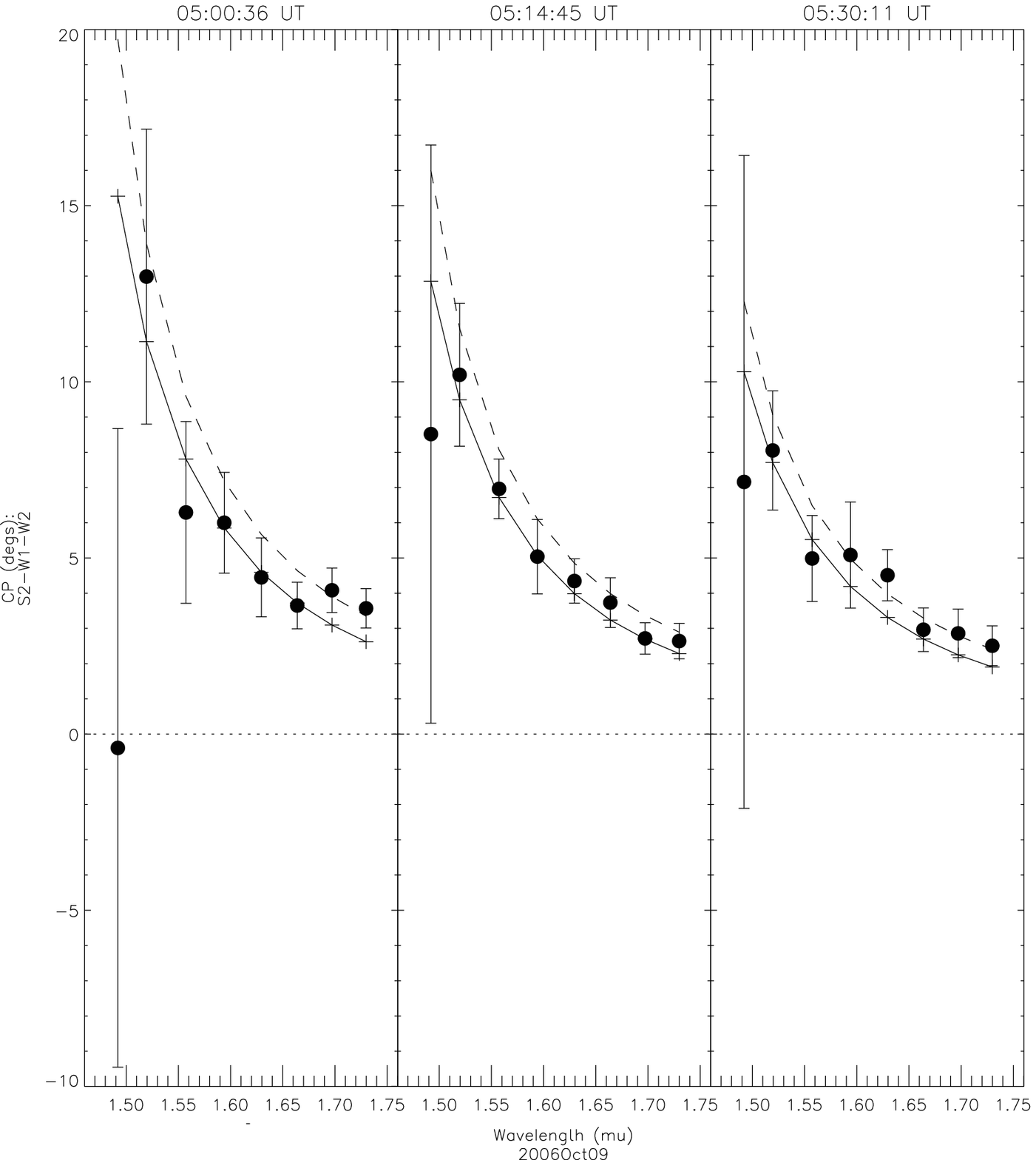}
\includegraphics[angle=0,width=3.2in]{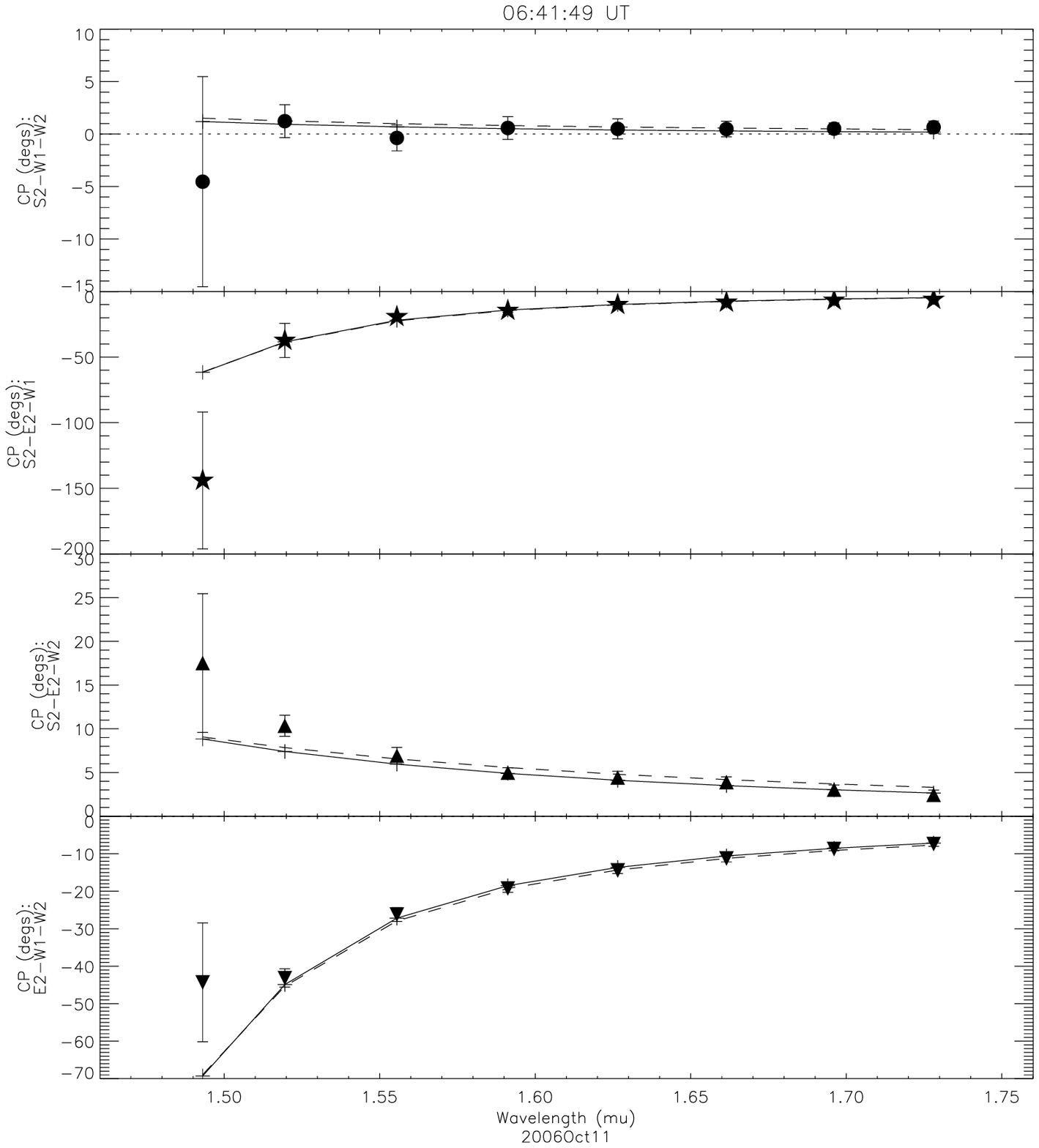}
\includegraphics[angle=0,width=3.2in]{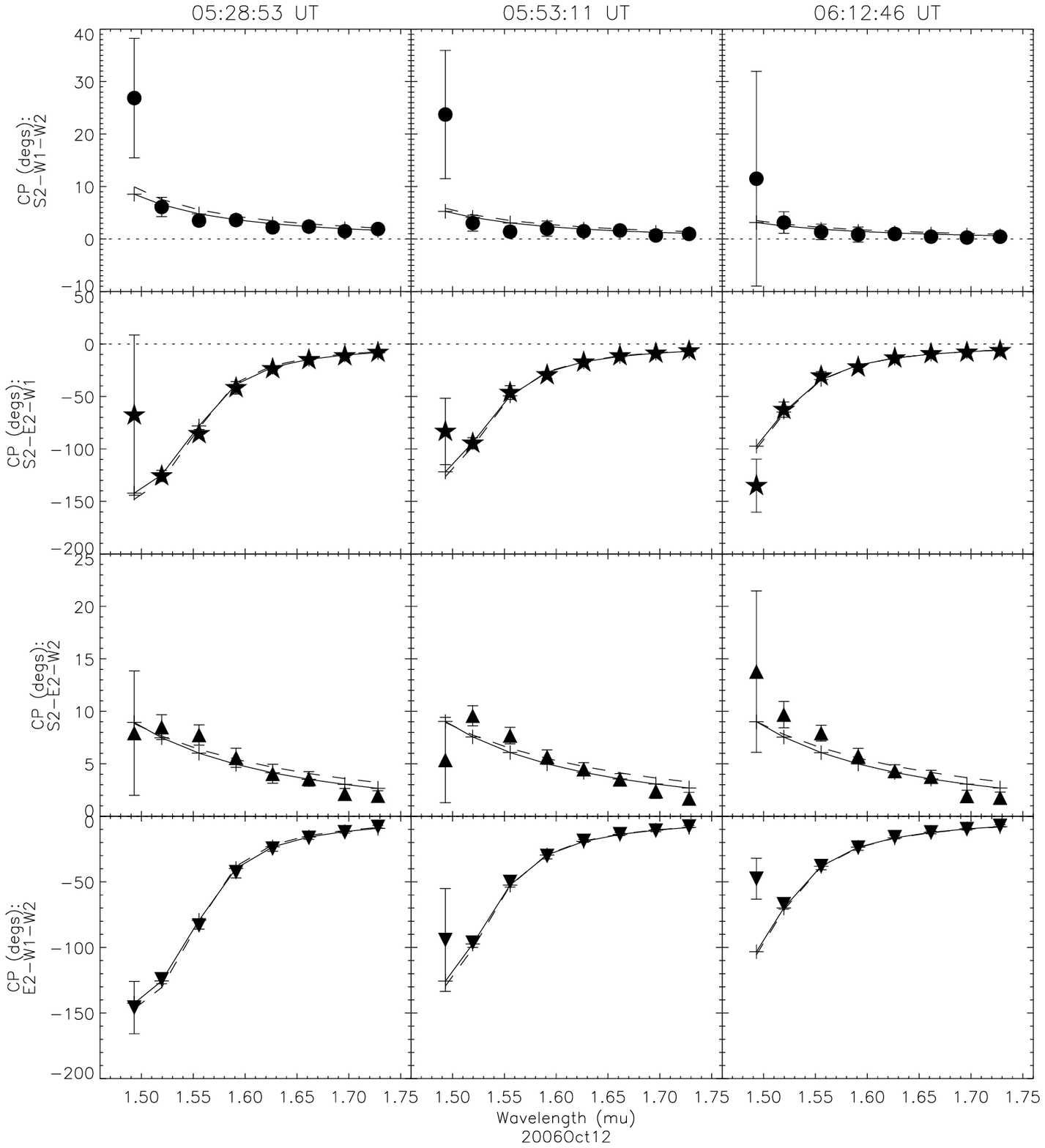}
\includegraphics[angle=0,width=3.2in]{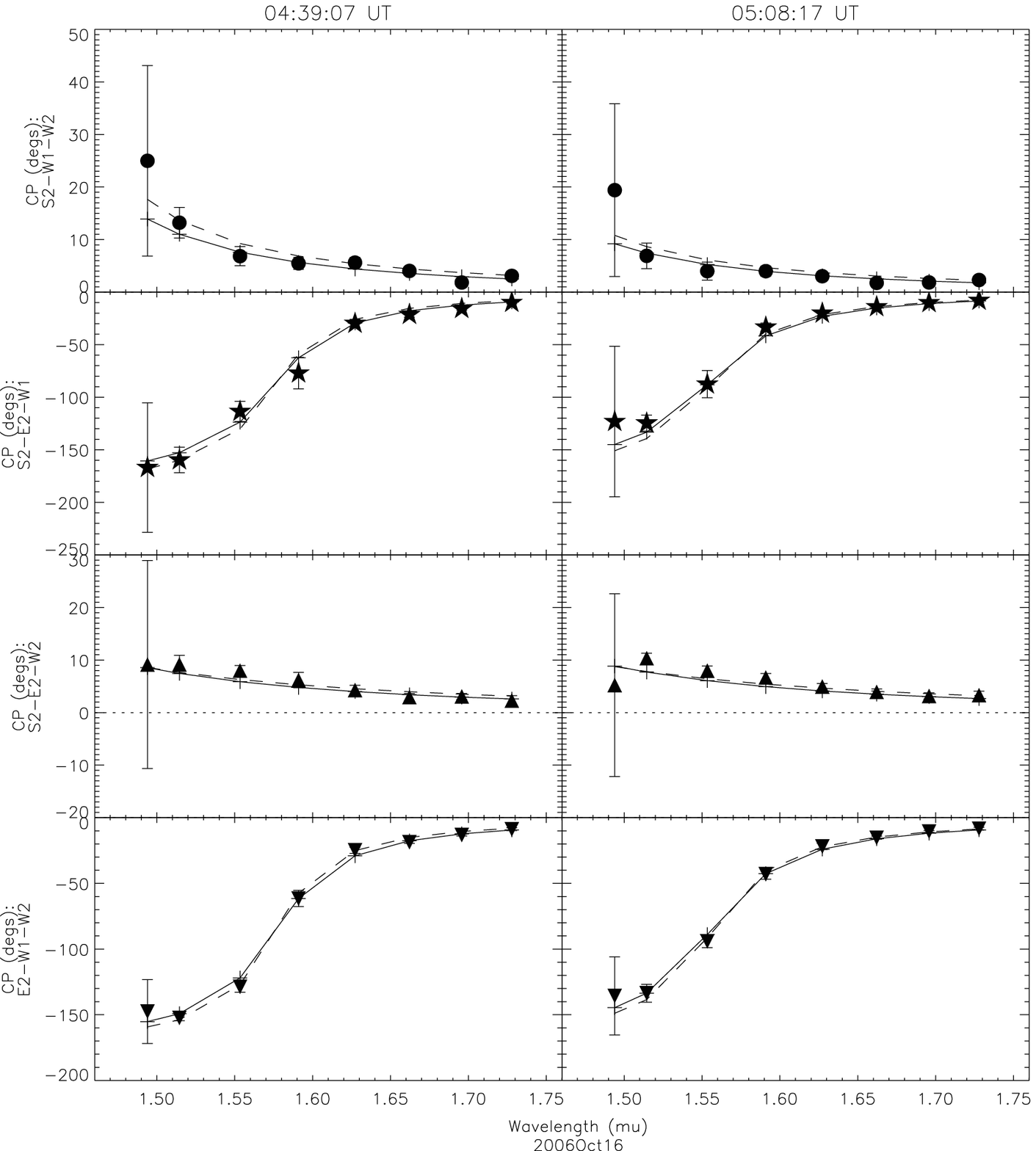}
}
\hphantom{.....}
\caption{ 
Similar to Fig.\ref{alfcep_vis2_img} but showing the closure phases for $\alpha$ Cep. The solid lines stand for the closure phases of the MACIM image, and the dashed lines stand for the model. Each row stands for a different telescope triangle. The $ \chi^2 _{\nu} $ of the image's closure phases is 0.95, while that of the model is 1.27.
\label{alfcep_cp_img}}
\end{center}
\end{figure}
\begin{figure}[thb]
{\includegraphics[angle=0,width=3.2in]{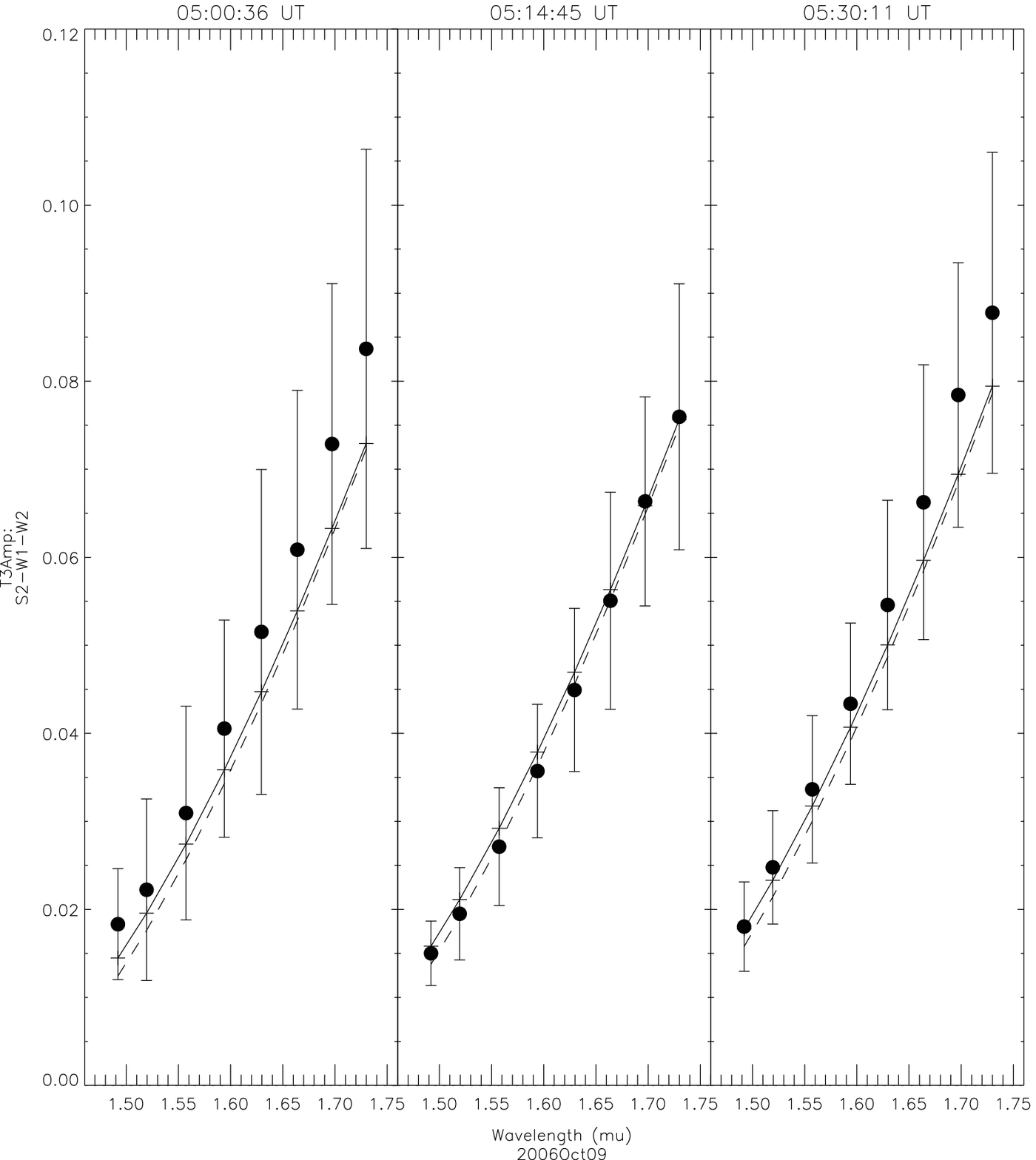}
\includegraphics[angle=0,width=3.3in]{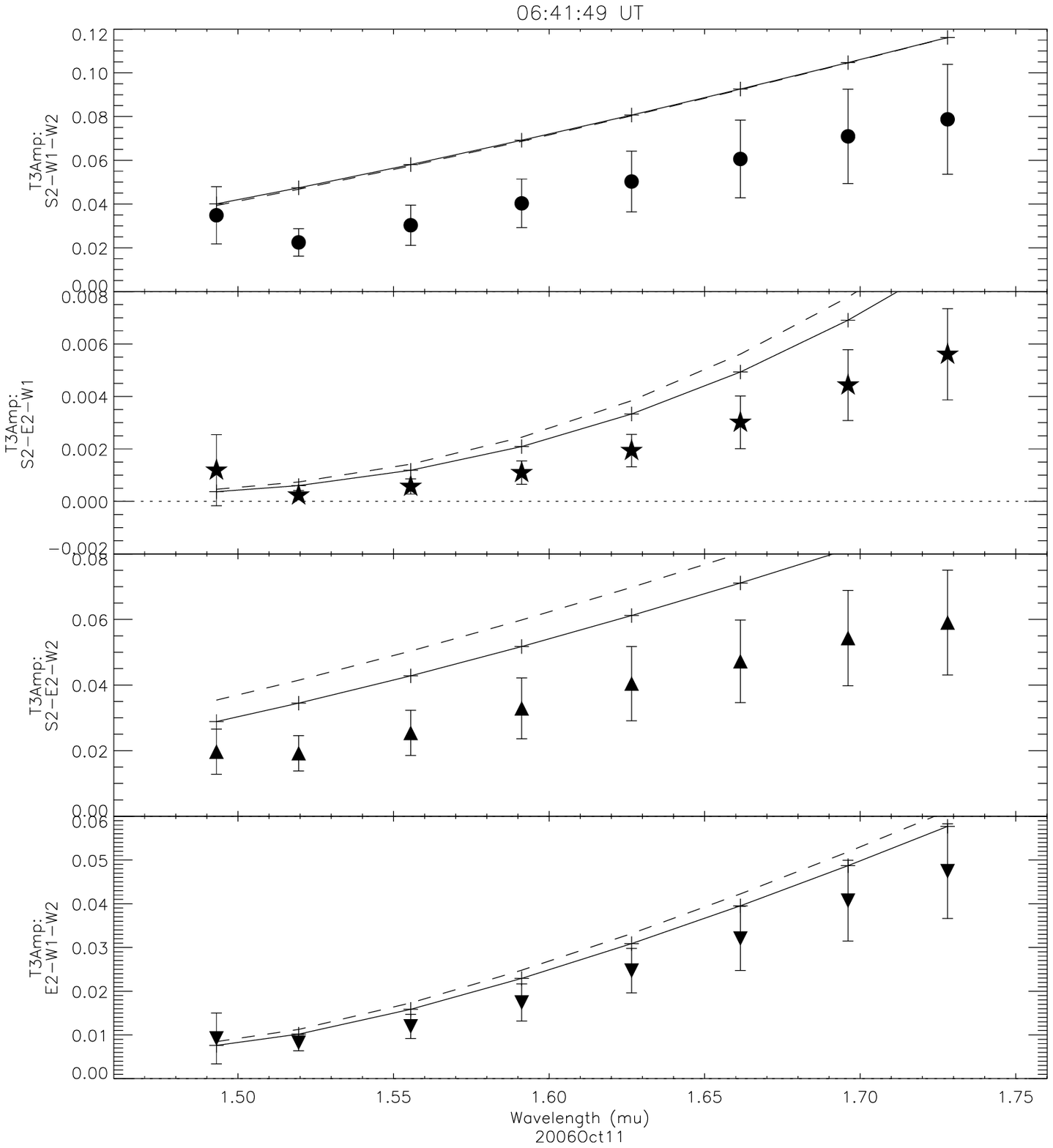}
\includegraphics[angle=0,width=3.2in]{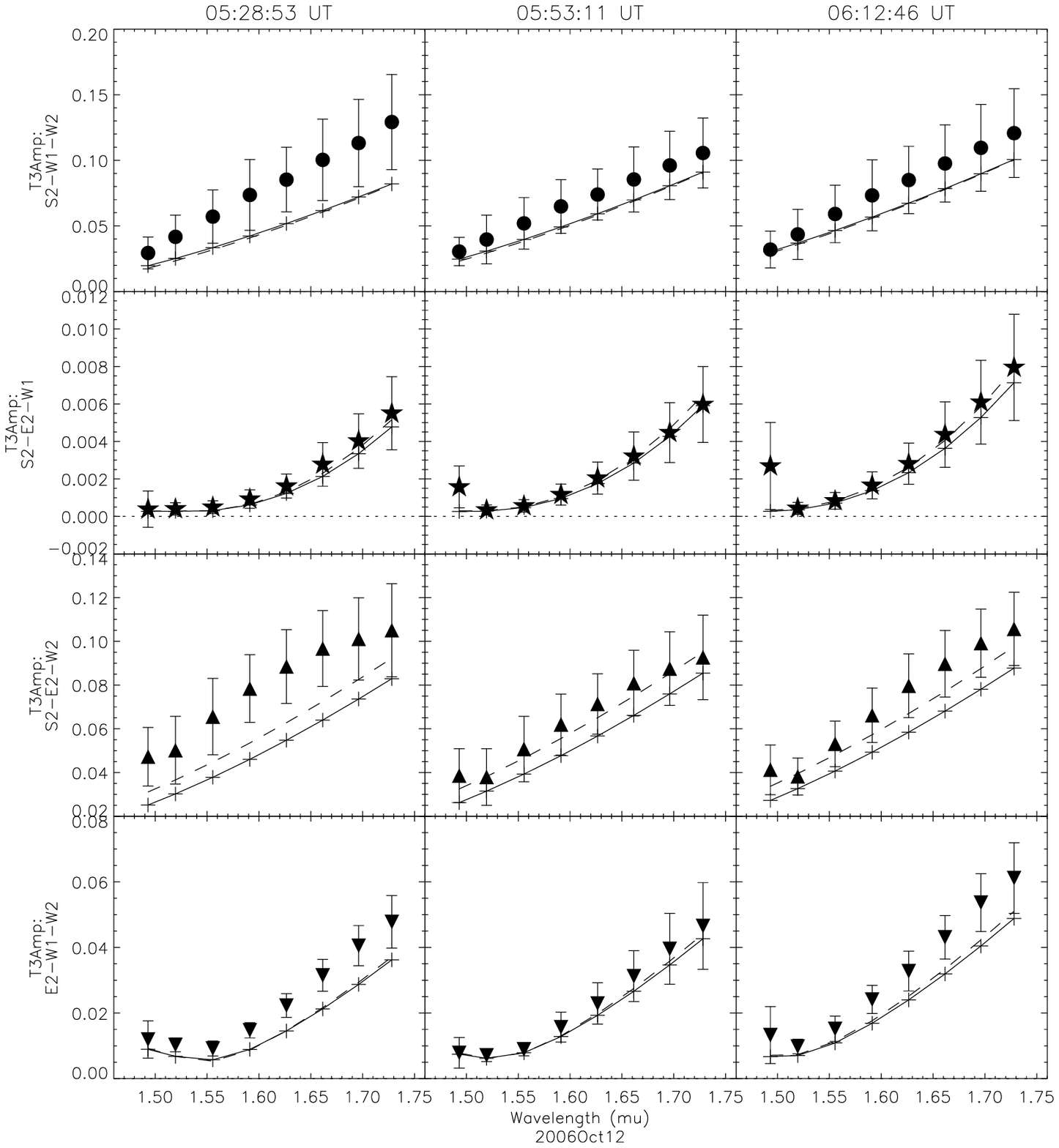}
\includegraphics[angle=0,width=3.2in]{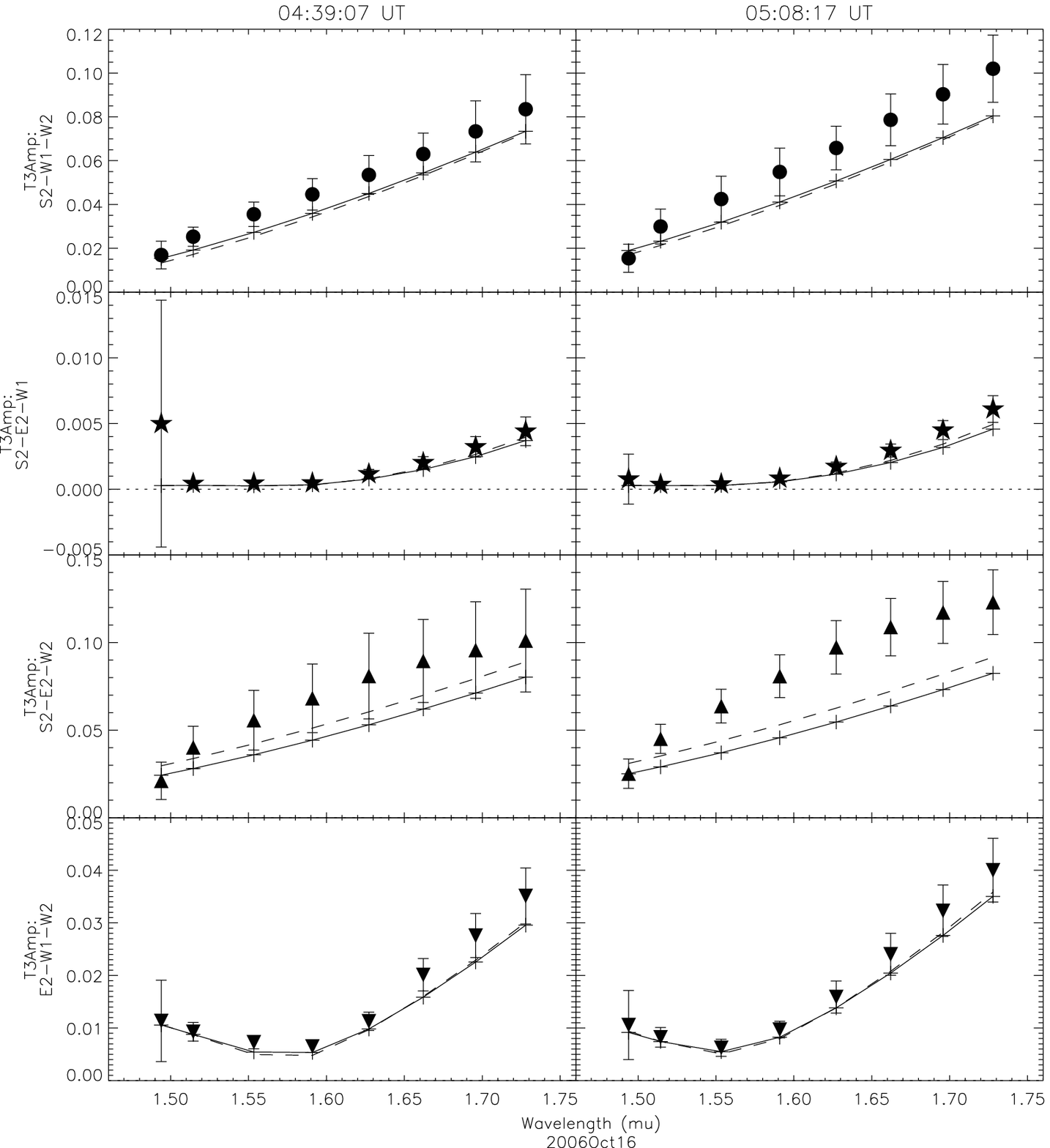}
}
\hphantom{.....}
\caption{ 
Similar to Fig.\ref{alfcep_vis2_img} but showing the triple amplitudes for $\alpha$ Cep. The $ \chi^2 _{\nu} $ of the image's triple amplitude is 1.63, while that of the model is 1.76.
 \label{alfcep_t3amp_img}}
\end{figure}

Figure \ref{alfcep_img} shows the reconstructed image of $\alpha$ Cep (${\chi_{\nu}}^2$ = 1.10). Its photosphere is well resolved and appears elongated along the east-west direction. 
The bright region at the bottom with T$_{eff}$ above 7000K (left panel) is later identified close to the pole and the dark belt below 6500K is the equator - a direct confirmation of the gravity darkening effect. 
The image implies the pole of \alfcep is medium inclined. The very top of the image becomes bright again since the photosphere is brighter toward the poles. The right panel of Fig.\ref{alfcep_img} shows the orientation of \alfcep based on the model in \S\ref{model}. It shows that the bright spot in the image is in fact above the pole as the pole of \alfcep is limb-darkened. The squared-visibilities, closure phases, and triple amplitudes derived from the image are compared with the data in Figure \ref{alfcep_vis2_img}, \ref{alfcep_cp_img}, and \ref{alfcep_t3amp_img}.

Although we have tried intensively to reconstruct an image for $\alpha$ Oph, we are unable to find a reliable solution for it. 
This is because the brightness distribution of a stellar surface is mainly imprinted in our closure phases. The closure phase is only sensitive to asymmetric structures of the object, while a symmetric object only gives either 0$^o$ or 180$^o$ closure phases. The squared-visibilities of our data are less constraining due to their relatively large errors.  The near equator-on inclination of \alfoph (see \S\ref{sec-alfoph}) makes its brightness distribution nearly symmetric, providing too few non-zero closure phase signatures to constrain the image. Therefore, we could not obtain a reliable solution for \alfoph in the image reconstruction. We have also pursued other imaging programs such as MIRA \citep{Thiebaut2008}, and obtained similar results in our preliminary efforts (Thi{\'e}baut 2008, private communication). Thus we only present the model of \alfoph in this paper. As we will see in \S\ref{sec-alfoph}, the lack of non-zero closure phase signatures of \alfoph also brings similar issues to our modeling, causing high degeneracy to the inclination and the gravity darkening coefficient.

\section{Surface Brightness Modeling}
\label{model}
In addition to synthesis imaging, we construct rapid rotator models to fit the data of both stars, following the prescription described in \citet{Aufdenberg2006} and references therein. Specifically, we assume a Roche potential (point mass) and solid body rotation in our model, and use the von Zeipel gravity darkening law \citep{vonZeipel1924, vonZeipel1924b} to characterize the latitudinal temperature profile. Six parameters are used to define the models, including the stellar radius and temperature at the pole, the angular rotation rate as a fraction of breakup ($\omega$), the gravity darkening coefficient ($\beta$), the inclination angle,  and the position angle (east of north)  of the star. To ensure accuracy of the models, we construct them at four different wavelength channels across the $H$ band. The intensity and limb darkening at each point of the stellar surface is interpolated using the stellar atmosphere models of Kurucz \citep{Kurucz1993} as a function of local temperature, gravity, viewing angle, and wavelength.  The 3D surfaces of the models are generated using patches with uniform surface areas to avoid over-sampling at the poles or under-sampling at the equators, and also to speed up the computation.
A direct Fourier transform is then used to convert the projected intensity model to squared-visibilities, closure phases and triple amplitudes \footnote{We have validated our model by comparing with another independent model from Jason Aufdenberg (private communication) on the data of Vega from \citet{Aufdenberg2006}. We also compared the model using Kurucz limb darkening with one using PHOENIX limb darkening and found the difference is negligible. The data and models we used for the comparison are available at http://www.astro.lsa.umich.edu/$\sim$mingzhao/rapidrot.php}.  In addition, we also force our model to match the $V$ and $H$ band photometric fluxes obtained from the literature (see Tables \ref{alfcep_tab}, \ref{alfoph_tab}) to constrain the temperature range.

\subsection{\alfcep}
\label{sec-alfcep}
\begin{figure}[thb]
\begin{center}
{\includegraphics[angle=90,width=3.2in]{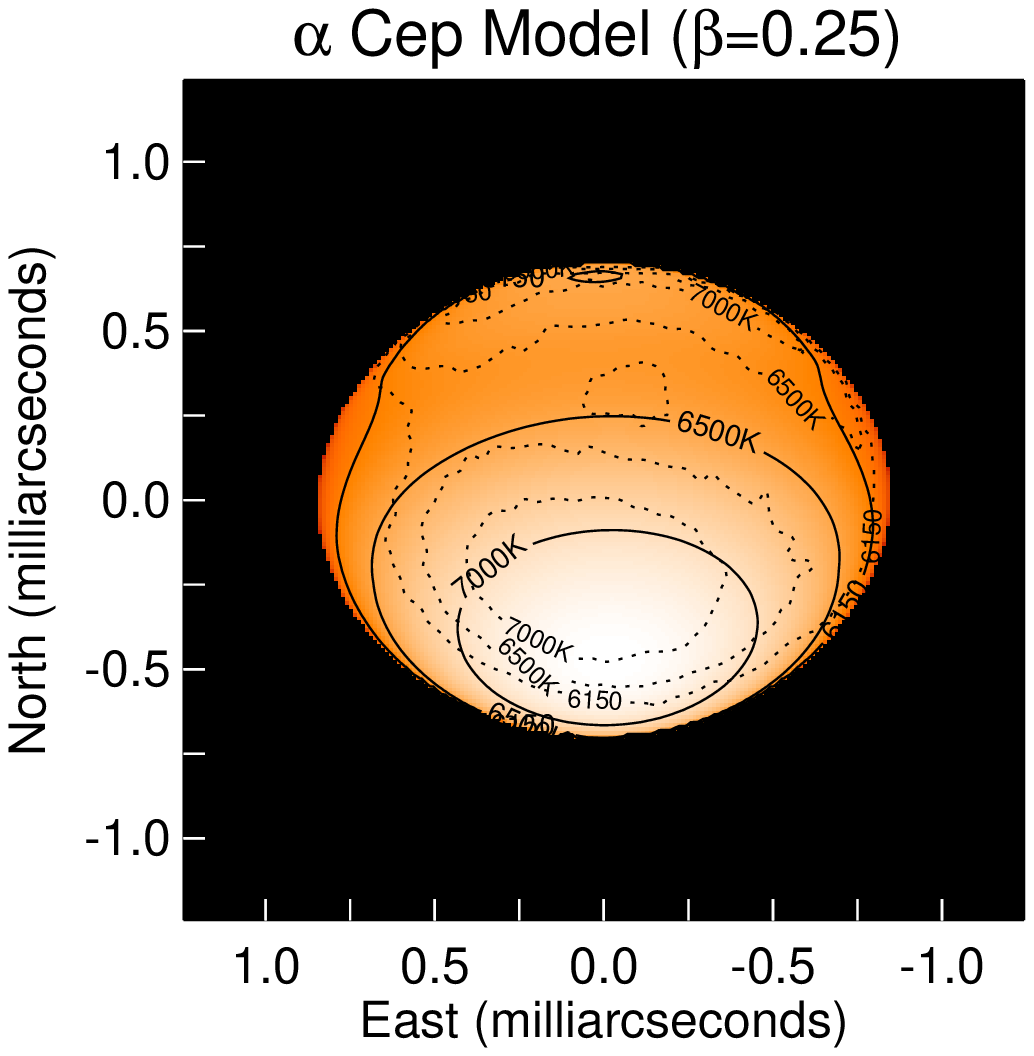}
\includegraphics[angle=90,width=3.2in]{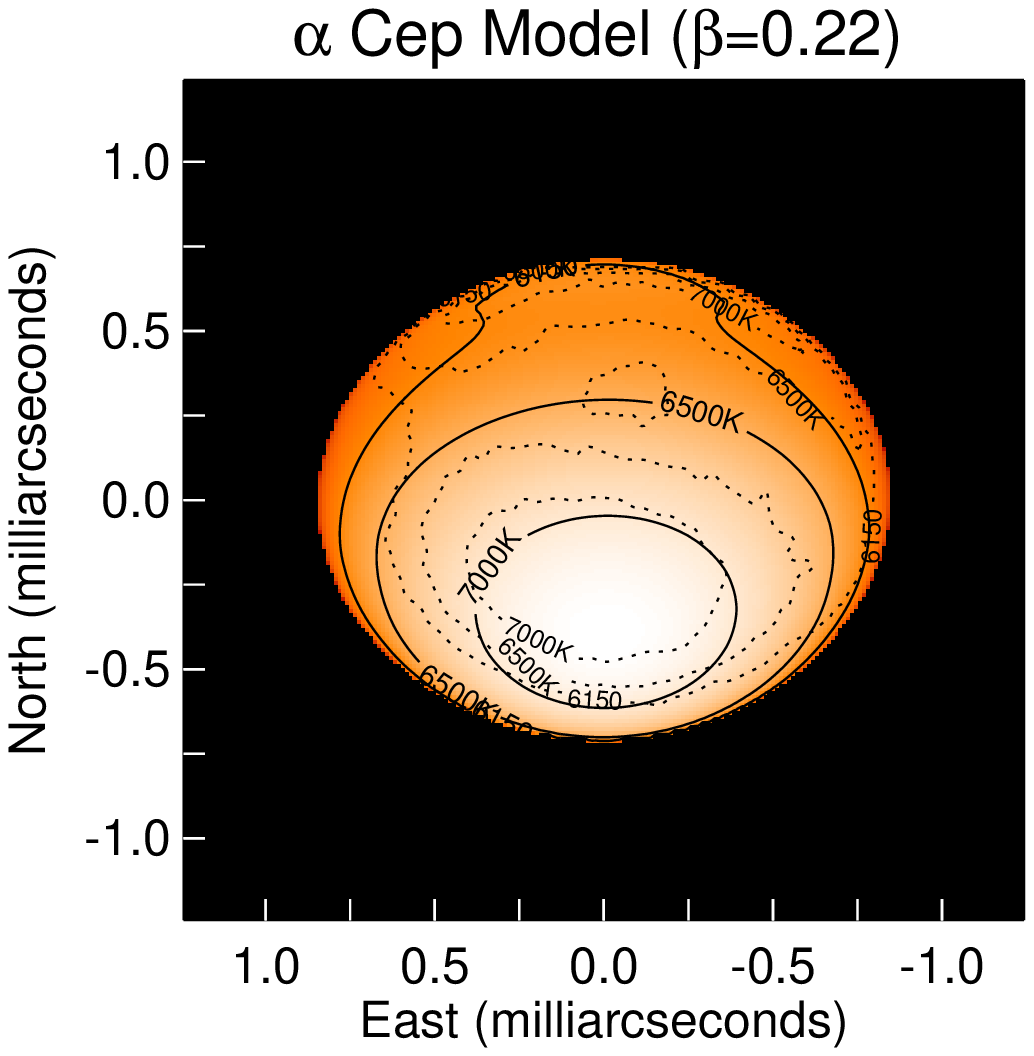}}
\hphantom{.....}
\caption{ 
The gravity darkening models of \alfcep. The contours indicate the local brightness temperatures on the surface of the star. The left panel shows the best-fit standard gravity darkening model ($\beta=0.25$) overplotted with the temperature 
 contours from Figure \ref{alfcep_img}. The total $\chi_{\nu}^2$ of the standard model is 1.21.  The right panel shows the best-fit $\beta$-free model, also overplotted with the temperature contours from  Figure \ref{alfcep_img},  and has a total $\chi_{\nu}^2$ of 1.18. The resolution of the data is 0.68 milliarcsec.
 \label{alfcep_model}}
\end{center}
\end{figure}

We first fit the data of \alfcep with the standard von Zeipel gravity darkening model for fully radiative envelopes (i.e., $T_{eff}\propto g_{eff}^{\beta}$, where $\beta=0.25$; hereafter, the standard model). The Levenberg-Marquardt  algorithm is applied for the least-square minimization and the parameter spaces are extensively searched in the fit. 
 We assume M = 2.0 \msun \citep{vanBelle2006},  distance = 14.96 pc \citep{Perryman1997}, and metallicity $[Fe/H]=0.09$ \citep{Gray2003} in the model. The left panel of Figure \ref{alfcep_model} shows the best-fit standard model of $\alpha$ Cep, with an overall goodness of fit $\chi_{\nu}^2$ of 1.21.  The model shows the photosphere of \alfcep is elongated, with a bright polar region at the bottom and a dark equator above it - generally consistent with the synthesized image in Fig.\ref{alfcep_img}.  Our standard model yields an inclination of $64\fdg9\pm4\fdg1$ and a position angle of $-178\fdg3\pm4\fdg1$, consistent with the ellipse fit of \citet[hereafter VB06]{vanBelle2006}, which gave a position angle of -177$^o$ (or 3$^o$ depending on the definition). However, both  the inclination and the position angle of their gravity darkening model ($i=88\fdg2, P.A.=17^o ~or~ -163^o$) differ from our results, as
we have better UV coverage and also closure phase information which is very sensitive to asymmetric structures. Our model indicates \alfcep is rotating very fast, at 92.6\% of its break-up speed. The temperature at the poles is $\sim$2400K higher than at the equator, while its radius at the equator is 26\% larger than at the poles. The best-fit parameters of the standard model are listed in the second column of Table \ref{alfcep_tab}. 
Since the calibration errors vary from night to night, we estimate the parameter
errors by bootstrapping the data from different nights (i.e., treat each night of data as a whole and 
 randomly sample all of the nights with replacement, so that the correlations of data within each 
night can be taken into account ) and fitting the parameters to the resampled data. 
We then iterated this procedure hundreds of times. 

In addition to our data, we also combine the squared-visibilities from VB06 (here after ``Classic data") into our fit. 
The combined fit gives a slightly higher inclination,  but all parameters are still consistent with our original fit. The total ${\chi_{\nu}}^2$ of the combined fit is 1.25. However, the  ${\chi_{\nu}}^2$ of the Classic data (${\chi_{\nu}}^2$=2.0) is  very large although it is slightly better than the original result of VB06 (${\chi_{\nu}}^2$ =2.16), implying that either the Classic data have additional un-calibrated errors or the model needs more degrees of freedom. We first look into a free $\beta$ in the model. Indeed, the von Zeipel theory suggests that the standard gravity darkening coefficient ($\beta=0.25$) only applies to pure radiative envelopes. However, it is uncertain if \alfcep is pure radiative or not. The atmosphere models of \citet{Kurucz1979} suggest that,  for an atmosphere with T$_{eff}>7500K$ and $\log g \sim 4$, like the polar areas of $\alpha$ Cep,  convection should have very little or no effect. But it starts to play a role when temperature and $\log g$ drop below those numbers. In addition,  the evolution models of $\beta$ calculated by \citet{Claret1998, Claret2000} also indicate that, for a 2-2.5\msun~star, convection starts to take place once T$_{eff}$ is below $\sim$7900K. For the case of $\alpha$ Cep,  although its T$_{eff}$s at the polar areas are higher than 8000K, they drop to only $\sim$6700K in the equator, implying that convection may have effects in the equatorial areas and $\beta$ may deviate from the standard value. Therefore as a preliminary effort, we extend the standard von Zeipel law to a free $\beta$.

 The new combined $\beta$-free fit gives a ${\chi_{\nu}}^2$ of 2.11 to the Classic data, similar to the original VB06's result. But it prefers a $\beta$ of 0.22 rather than the 0.08 value of VB06. To address this issue, we tried to fit the combined data at a fixed $\beta$ of 0.08 instead, but only obtained a total ${\chi_{\nu}}^2$ of  $\sim6.5$, much worse than the previous result. 
In addition, we also fit the Classic data only but found it is too hard to constrain the model due to the small amount of data and lack of phase information.
Therefore, due to possible uncertainties of the Classic data, we applied the $\beta$-free model to the MIRC measurements only, and the results are shown in the third column of  Table \ref{alfcep_tab}. The best-fit model is shown in the right panel of Figure \ref{alfcep_model}. The squared-visibilities, closure phases and triple amplitudes of the $\beta$-free model are compared with the data in Figures \ref{alfcep_vis2_img}, \ref{alfcep_cp_img}, and \ref{alfcep_t3amp_img}, respectively. 

\begin{figure}[thb]
\begin{center}
{\includegraphics[angle=90,width=3.4in]{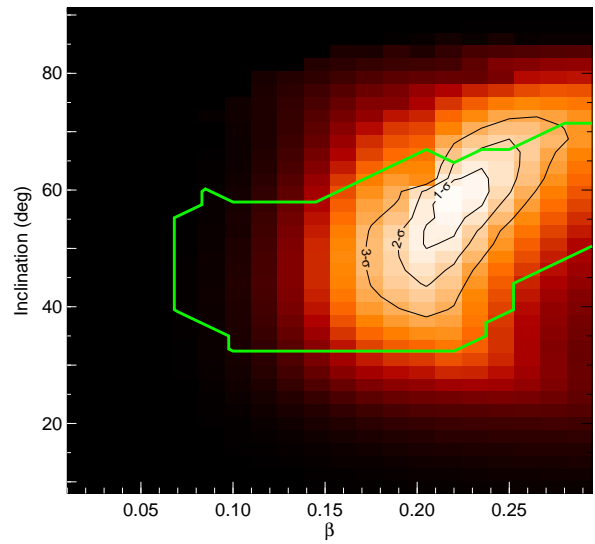}
}
\hphantom{.....}
\caption{ 
The $\chi_{\nu}^2$ surface of $\beta$ and inclination for $\alpha$ Cep. The corresponding probability peaks at $\beta\sim0.22$ and $i\sim56^o$. The black contours show the 1-$\sigma$, 2-$\sigma$, and 3-$\sigma$ levels of confidence interval, scaled to match the errors of $\beta$ and inclination estimated from bootstraping. The area inside the green box indicates the region where the corresponding V$sin$i values are within the observed range of 180 - 245 \kms.
 \label{alpcep_i_b}}
\end{center}
\end{figure}

The right panel of Figure \ref{alfcep_model} shows that the $\beta$-free model is more consistent with the synthesized image in Fig.\ref{alfcep_img} than the standard model. 
The ${\chi_{\nu}}^2$  of closure phase is significantly improved in the new best-fit although the ${\chi_{\nu}}^2$ of the triple amplitude is slightly larger.  
Figure \ref{alpcep_i_b}  illustrates the $\chi_{\nu}^2$ space of inclination and $\beta$ for $\alpha$ Cep, showing the value of $\beta$  is well constrained in the new model and is slightly lower than the standard value of 0.25. We also test the corresponding V$sin$i of the models in Fig.\ref{alpcep_i_b}. The peak of the $\chi_{\nu}^2$ space falls inside the green box, consistent with the observed range of V$sin$i.
The new model prefers a lower inclination of $55\fdg70\pm6\fdg23$, a higher rotational speed of 94$\%$ of break-up, and a similar position angle. The new best-fit temperatures at the poles and the equator are both cooler than those of the previous standard model.

In addition to using an average $\beta$ throughout the stellar surface as applied above, we are also pursuing fitting $\beta$ as a function of latitude.  This approach will be presented in a future work  with higher resolution data. 


\subsection{\alfoph}
\label{sec-alfoph}
\begin{figure}[thb]
\begin{center}
{
\includegraphics[angle=90,width=3.2in]{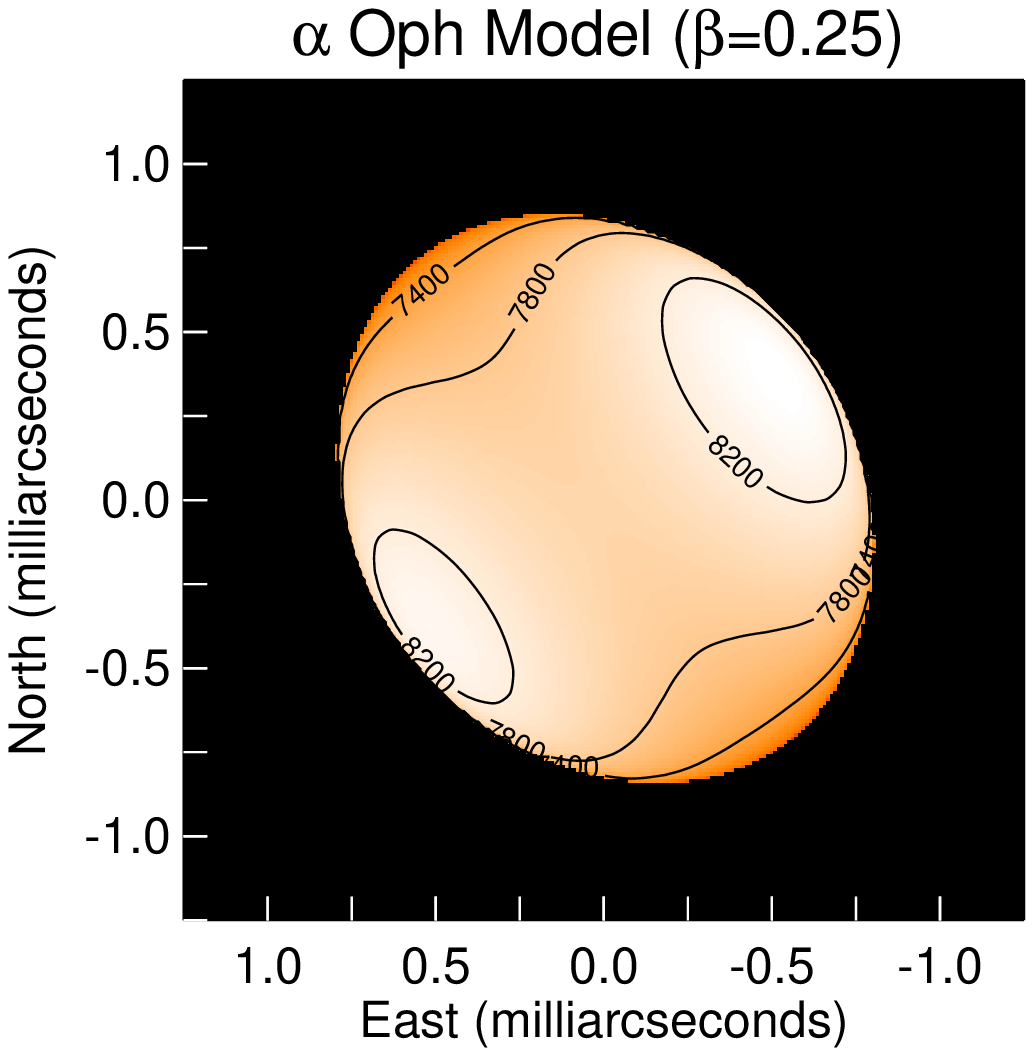}
\includegraphics[angle=90,width=3.2in]{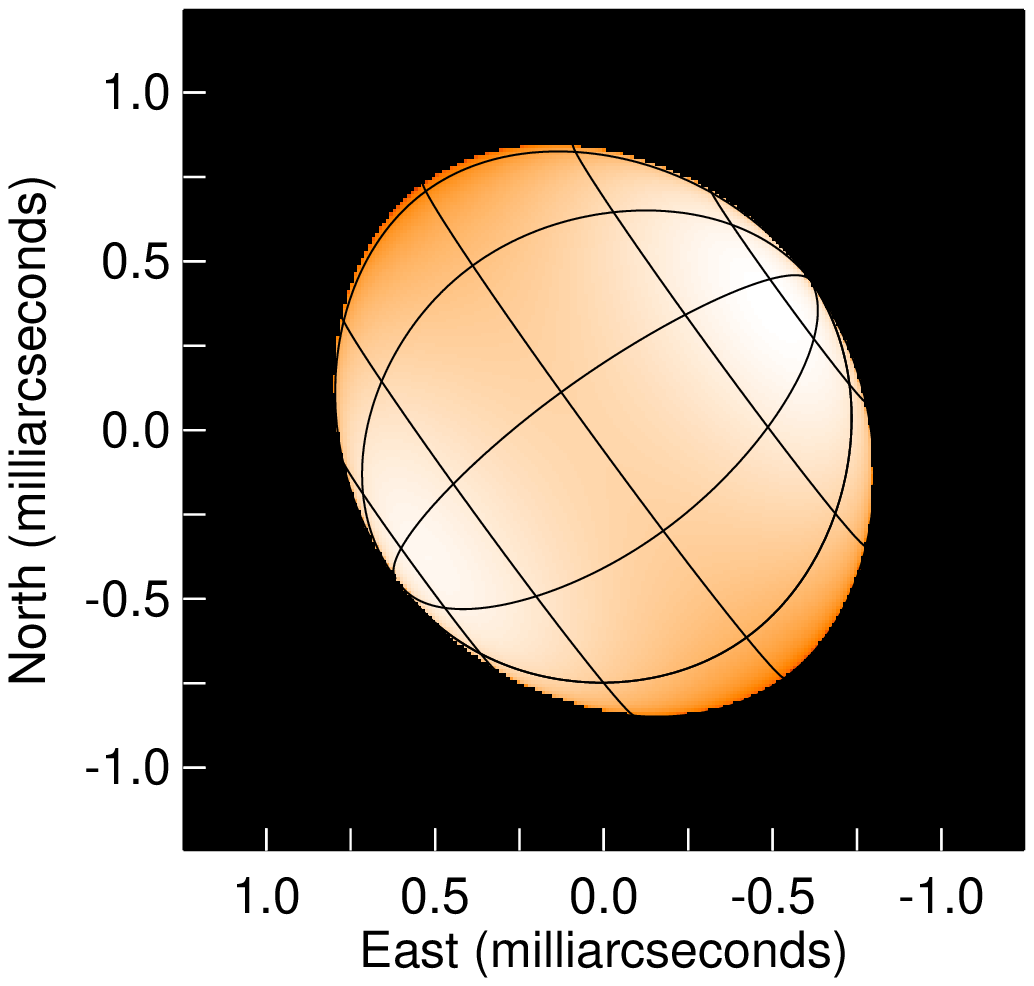}}
\hphantom{.....}
\caption{ 
The best-fit standard gravity darkening model of $\alpha$ Oph. The contours in the left panel indicate the local brightness temperatures on the surface of the star. The right panel shows the latitude and longitude of \alfoph to help visualize its geometry. The resolution of the data is 0.52 milliarcsec. The total $\chi_{\nu}^2$ of the model is 0.91.
\label{alfoph_model}}
\end{center}
\end{figure}

\begin{figure}[thb]
\begin{center}
{\includegraphics[angle=0,width=3.2in]{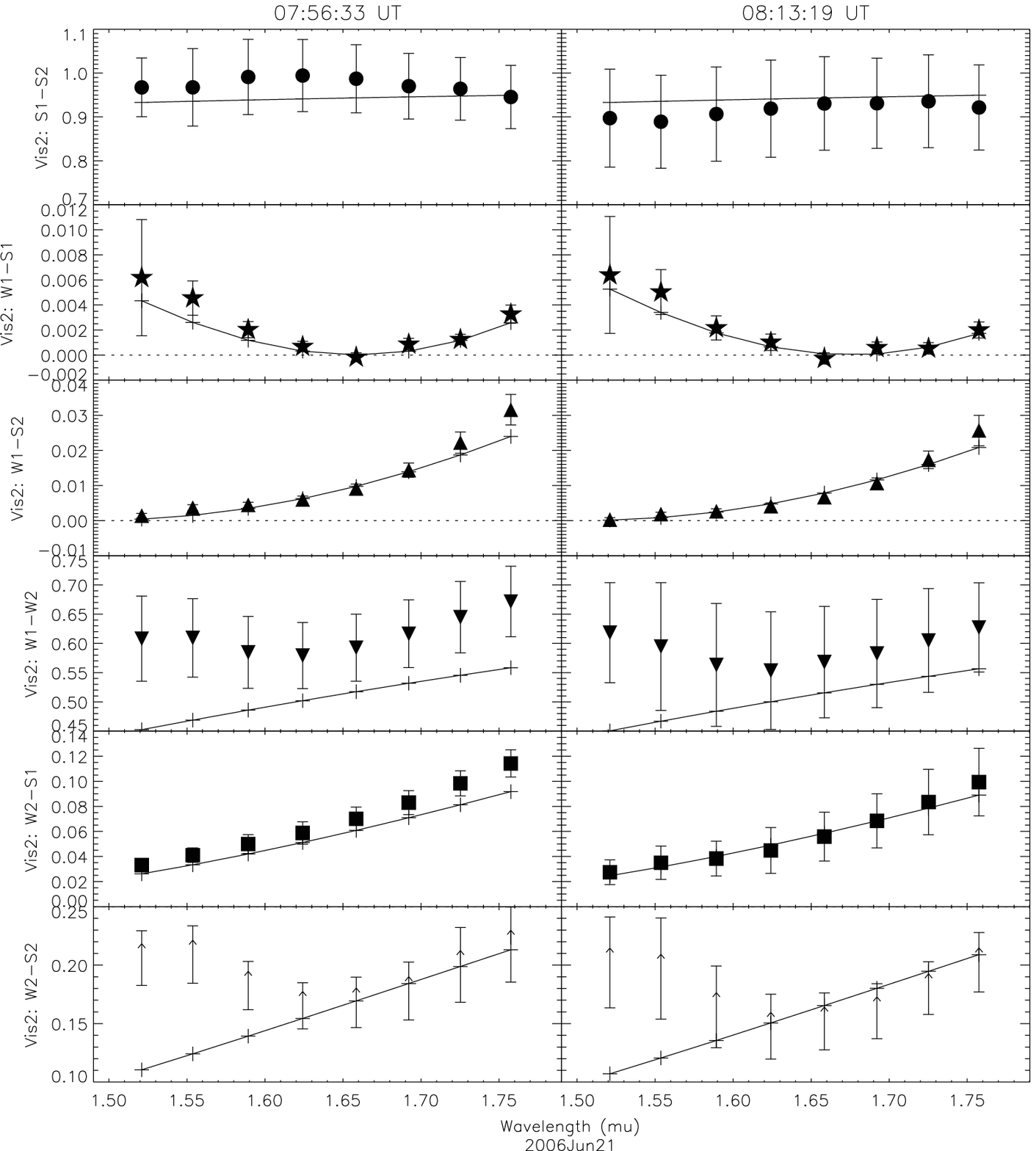}
\includegraphics[angle=0,width=3.2in]{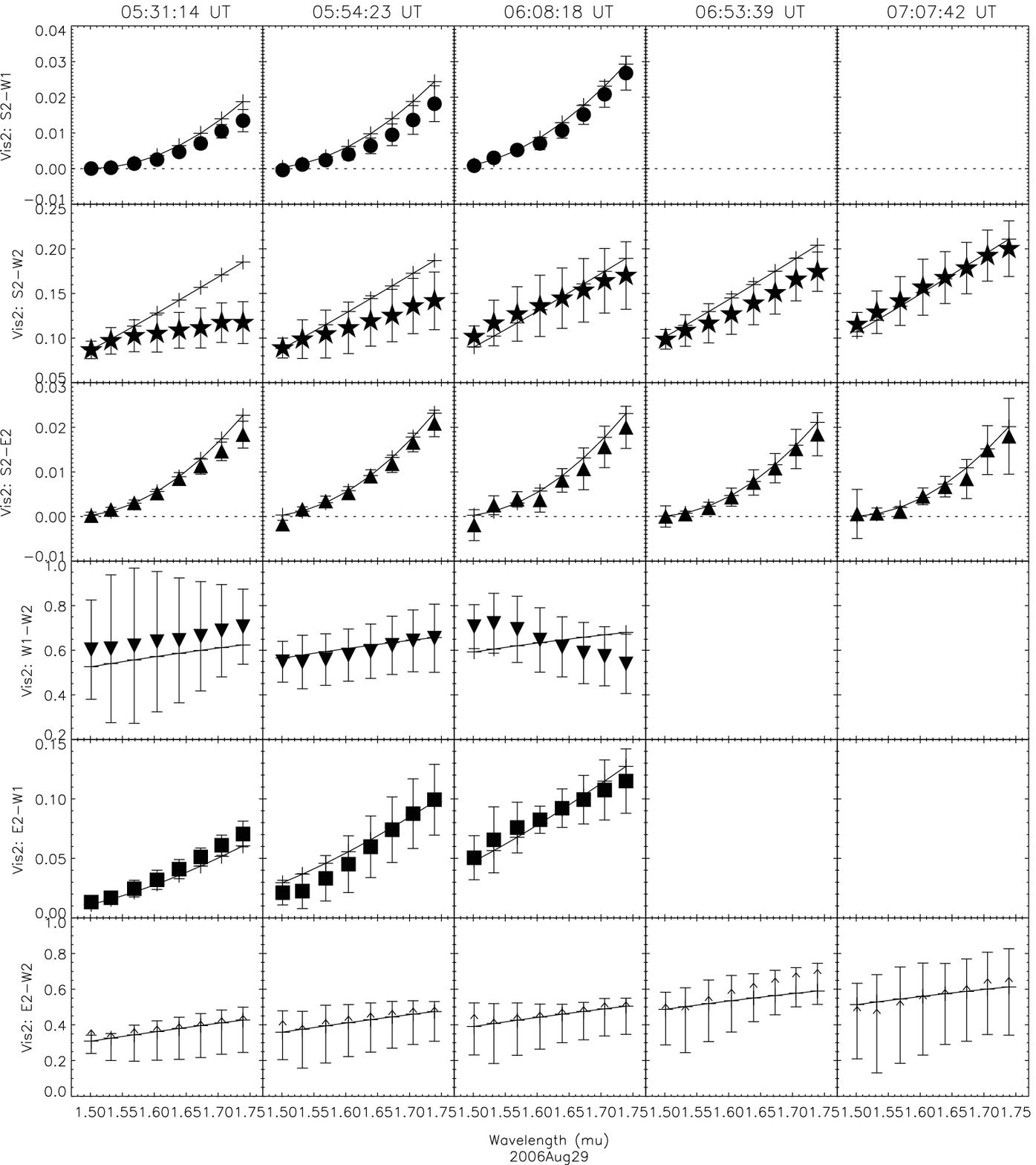}
\includegraphics[angle=0,width=3.2in]{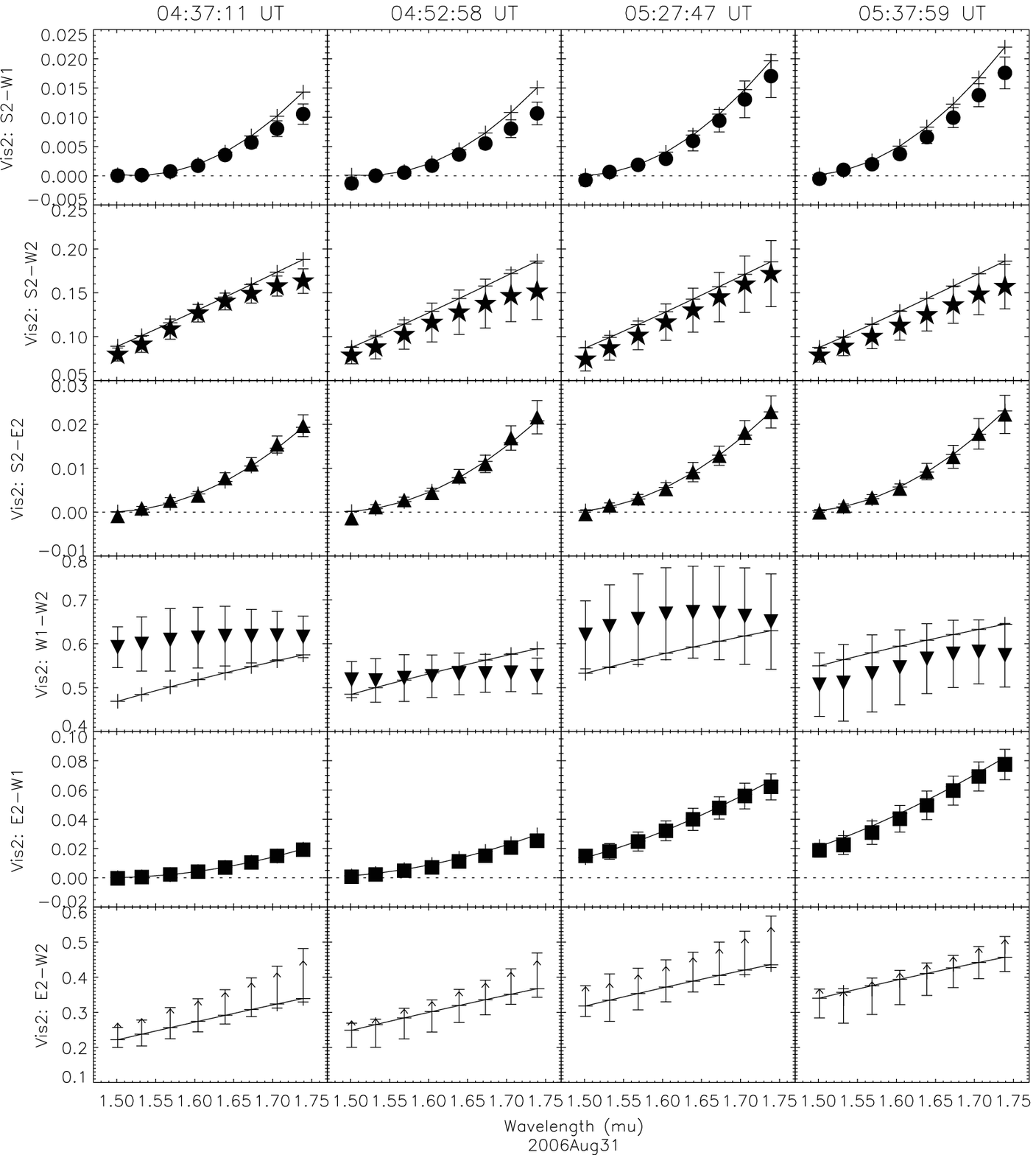}
\includegraphics[angle=0,width=3.2in]{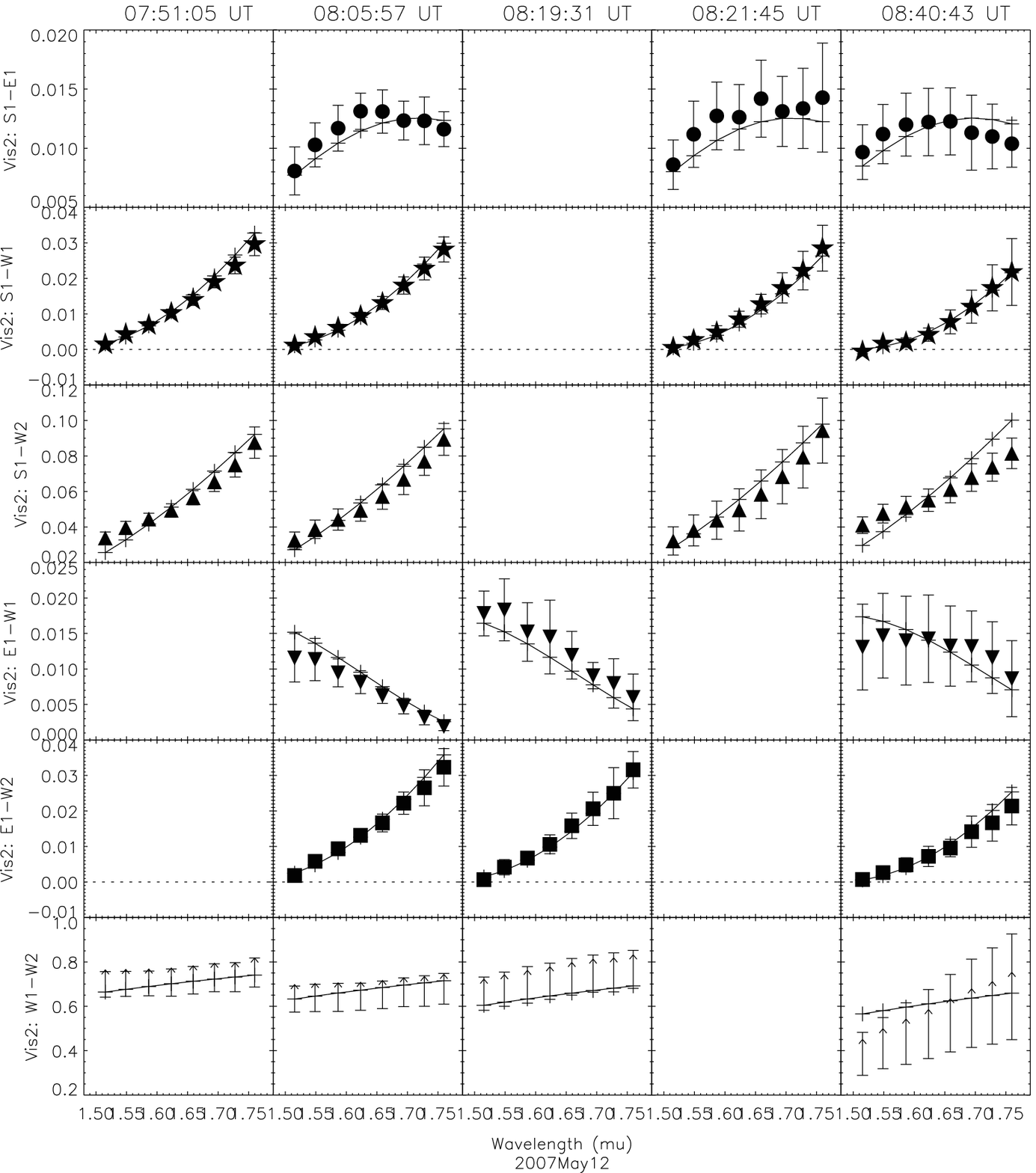}
}
\hphantom{.....}
\caption{
\alfoph squared-visibility model (standard $\beta=0.25$, solid lines) vs. data (filled points with error bars). Four out of eight nights (2006Jun21, 2006Aug29, 31, and 2007May12) are shown here.
Each row stands for a different baseline, while the columns indicate different times of observation. The eight data points in each panel indicate the eight spectral channels 
of MIRC across the $H$ band. The total $ \chi^2 _{\nu} $ is 0.72 for the squared-visibility only. (Please refer to the electronic edition if the type size is too small.)
 \label{alfoph_vis2}}
\end{center}
\end{figure}

\begin{figure}[thb]
\begin{center}
{\includegraphics[angle=0,width=3.2in, height= 3.8in]{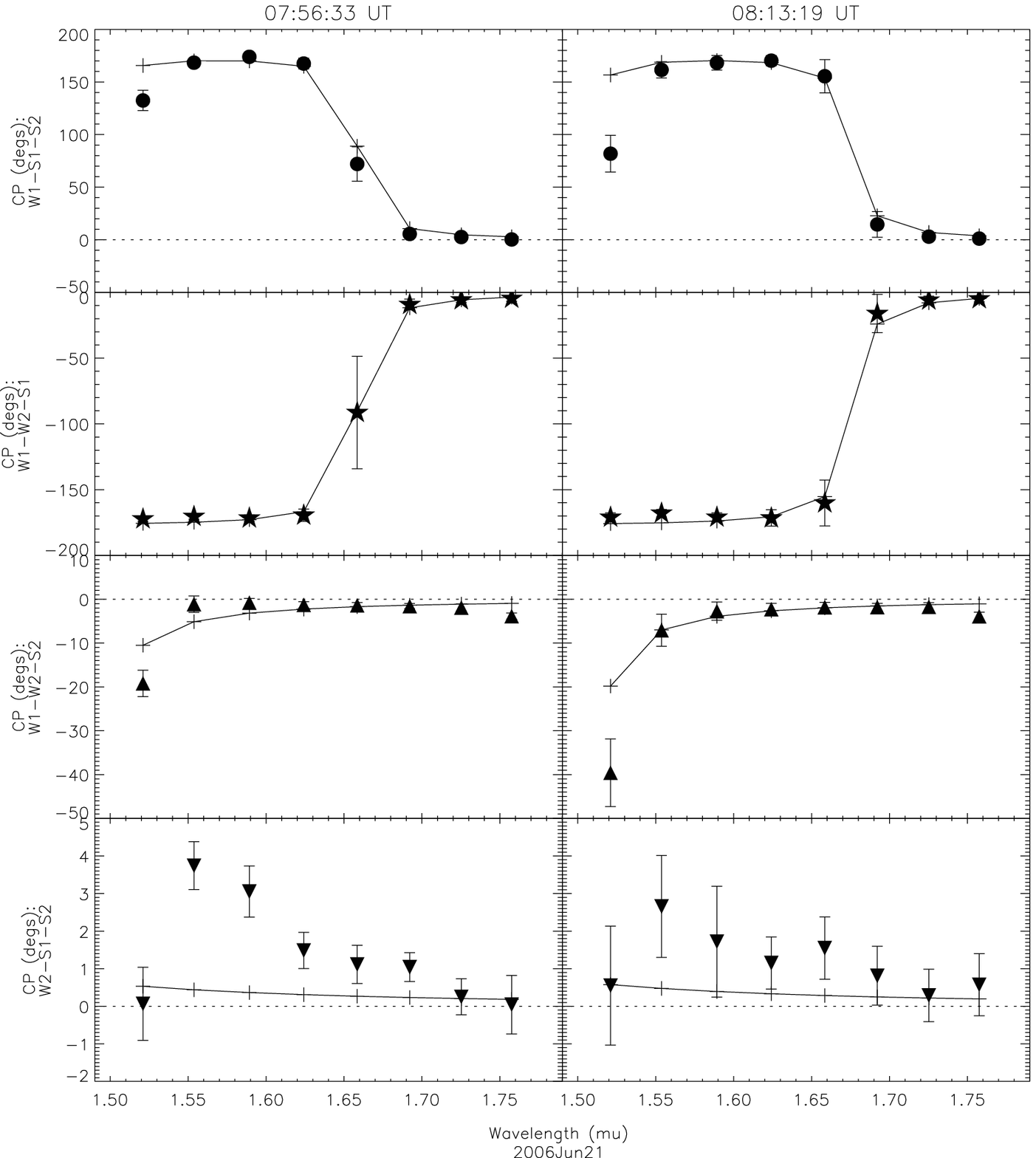}
\includegraphics[angle=0,width=3.2in,height= 3.8in]{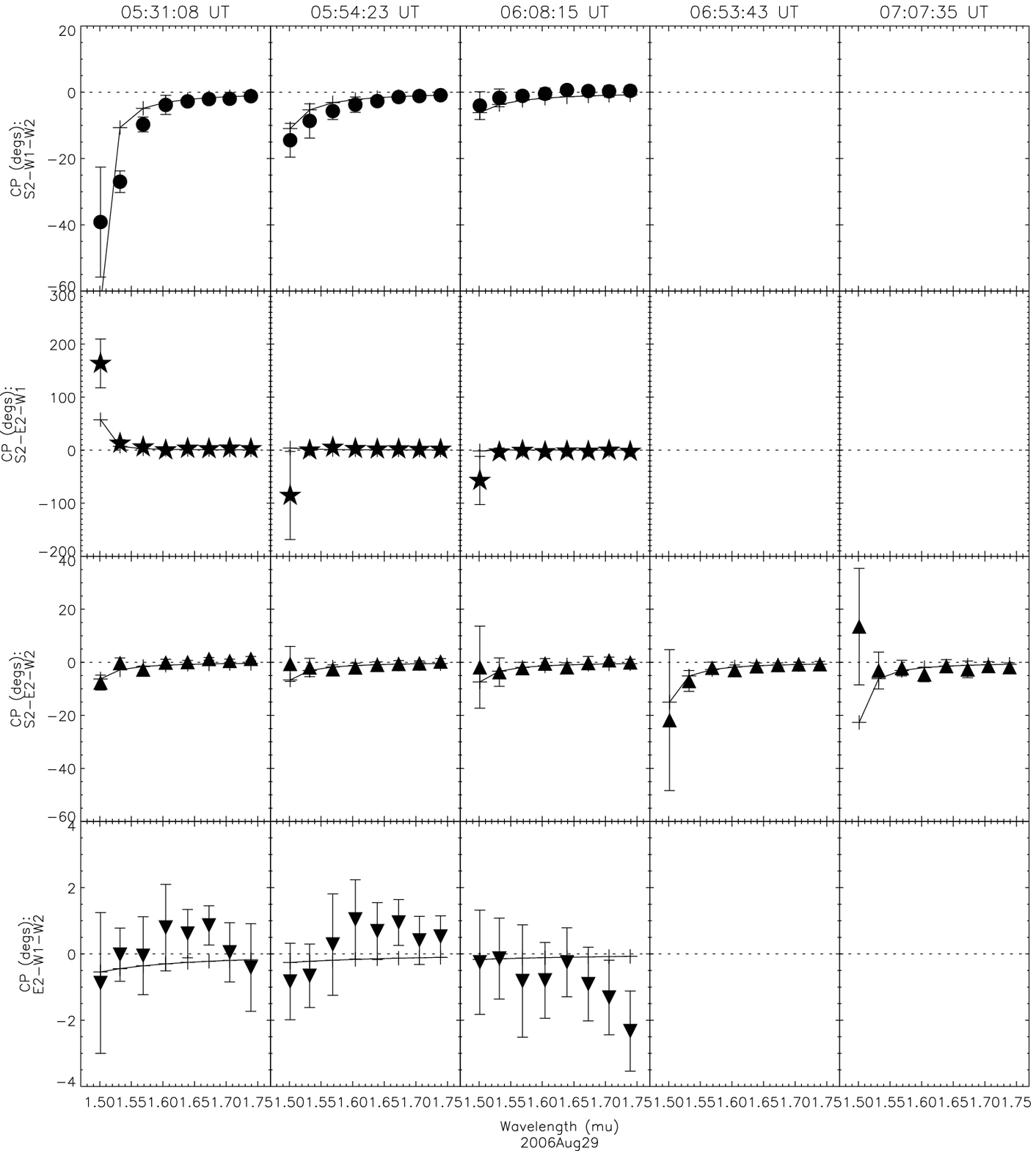}
\includegraphics[angle=0,width=3.2in, height=3.8in]{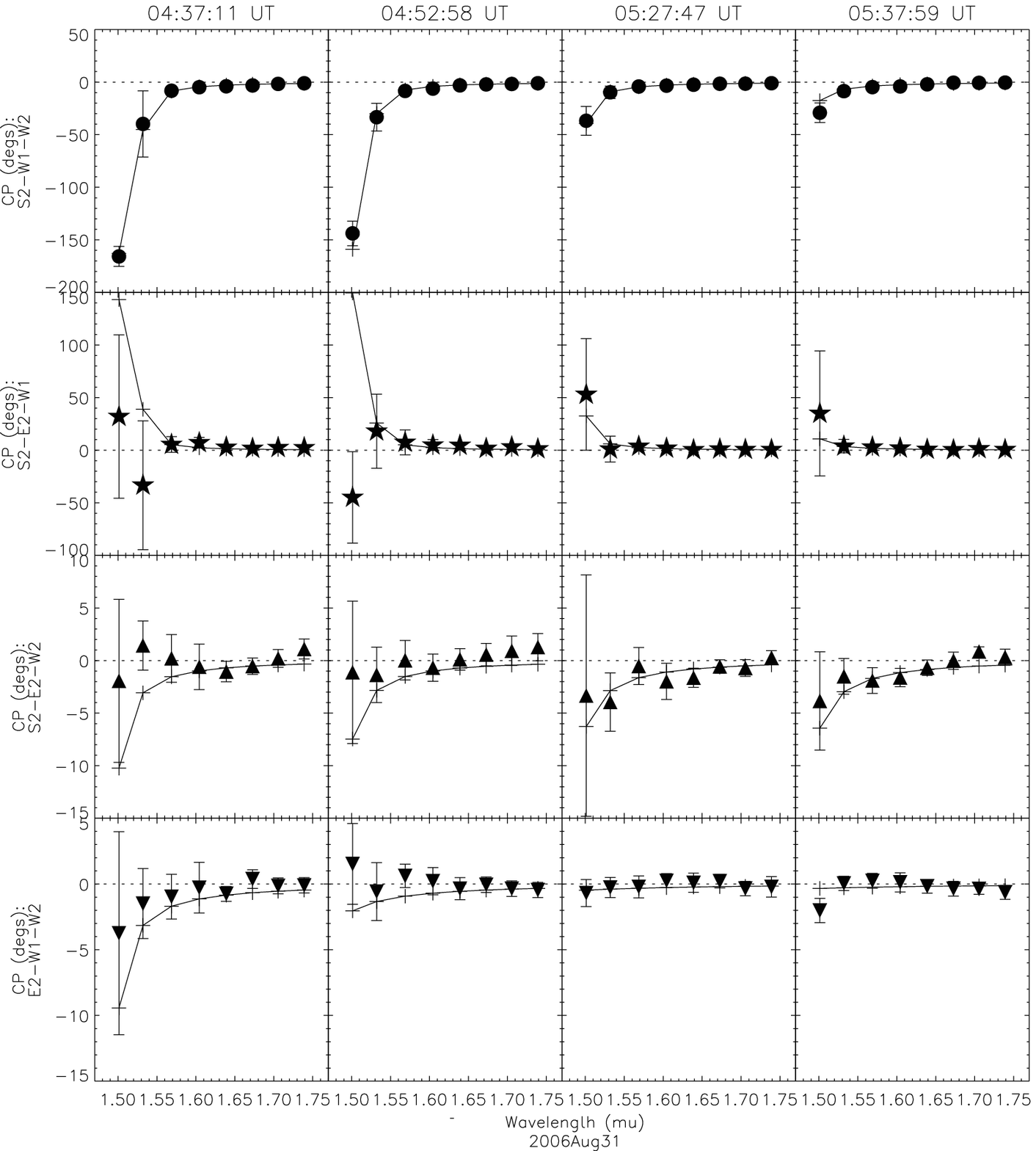}
\includegraphics[angle=0,width=3.2in, height= 3.8in]{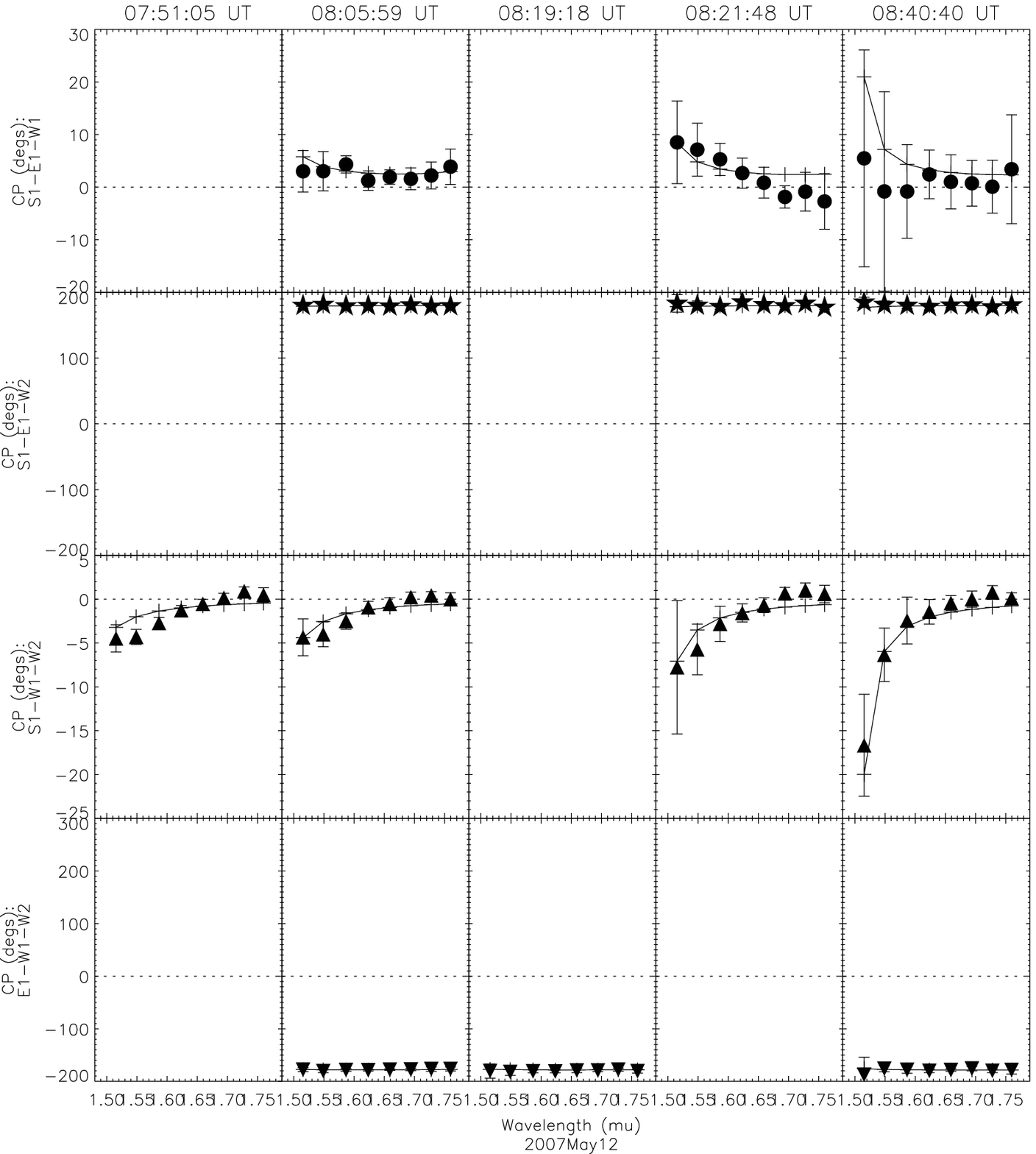}
}
\hphantom{.....}
\caption{ 
Similar to  Fig.\ref{alfoph_vis2} but showing the closure phase for $\alpha$ Oph. Each row stands for a different telescope triangle. The total $ \chi^2 _{\nu} $ for closure phase is 1.33.
 \label{alfoph_cp}}
\end{center}
\end{figure}

\begin{figure}[thb]
\begin{center}
{\includegraphics[angle=0,width=3.2in, height= 3.8in]{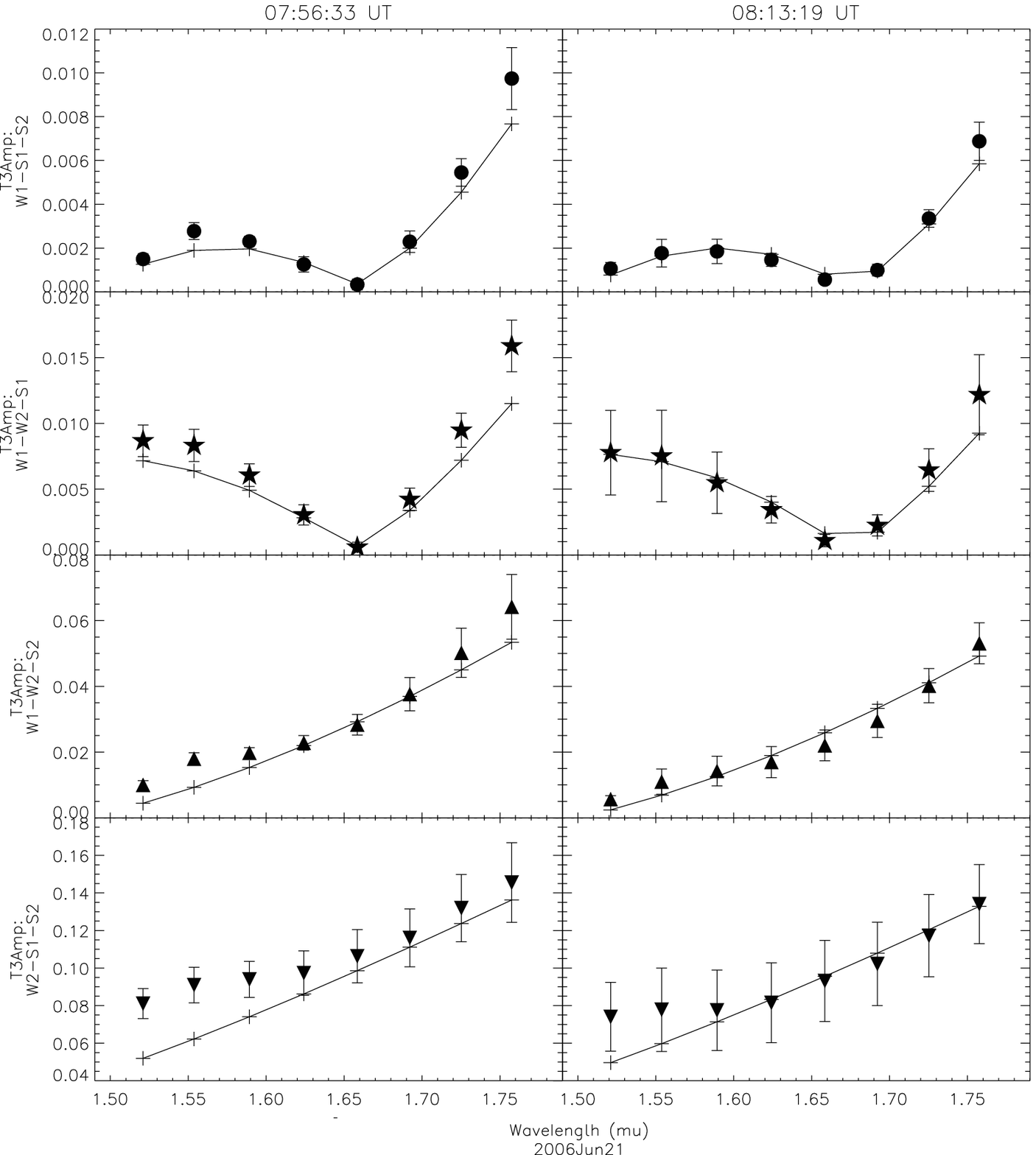}
\includegraphics[angle=0,width=3.2in,height= 3.8in]{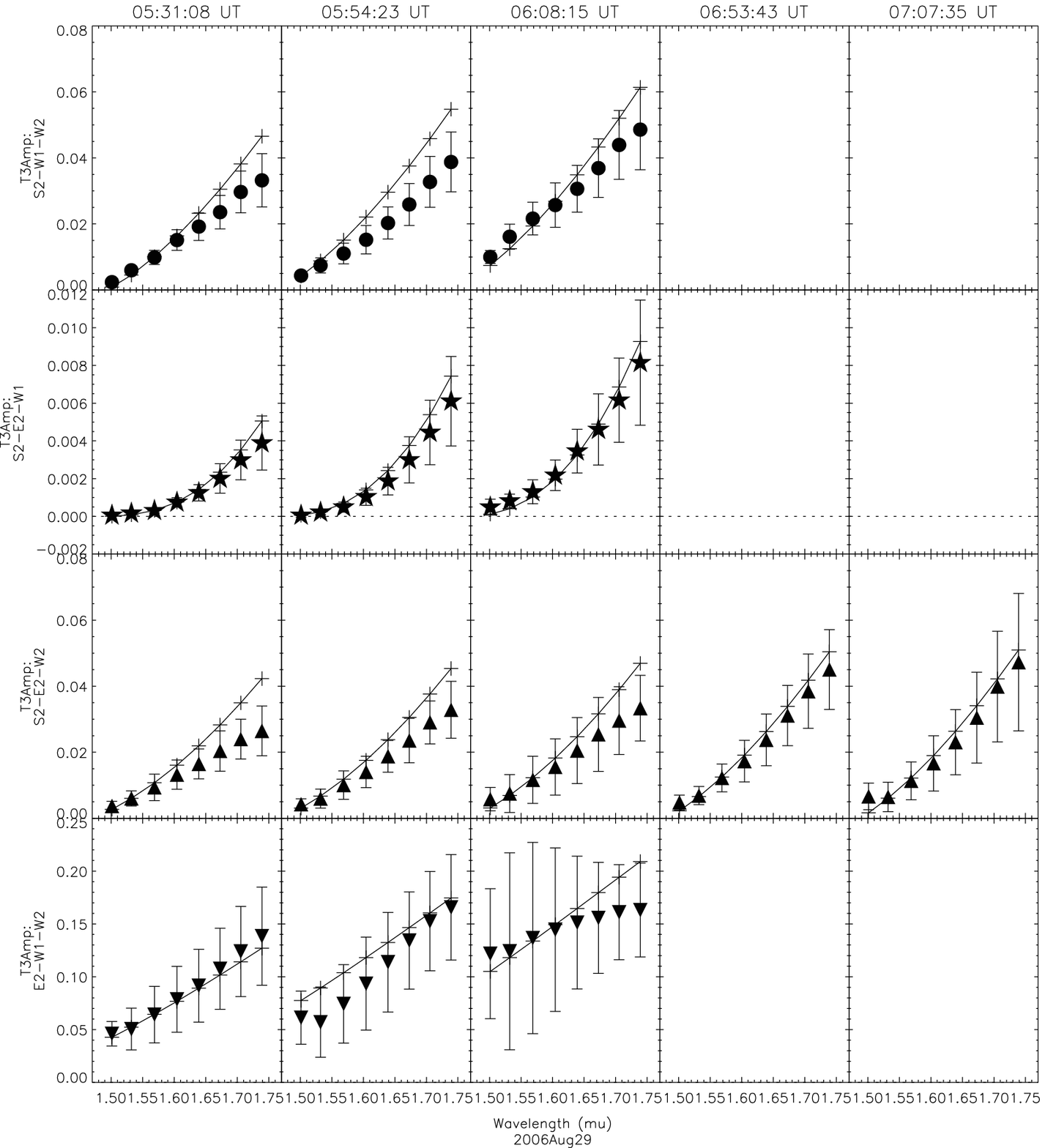}
\includegraphics[angle=0,width=3.2in, height=3.8in]{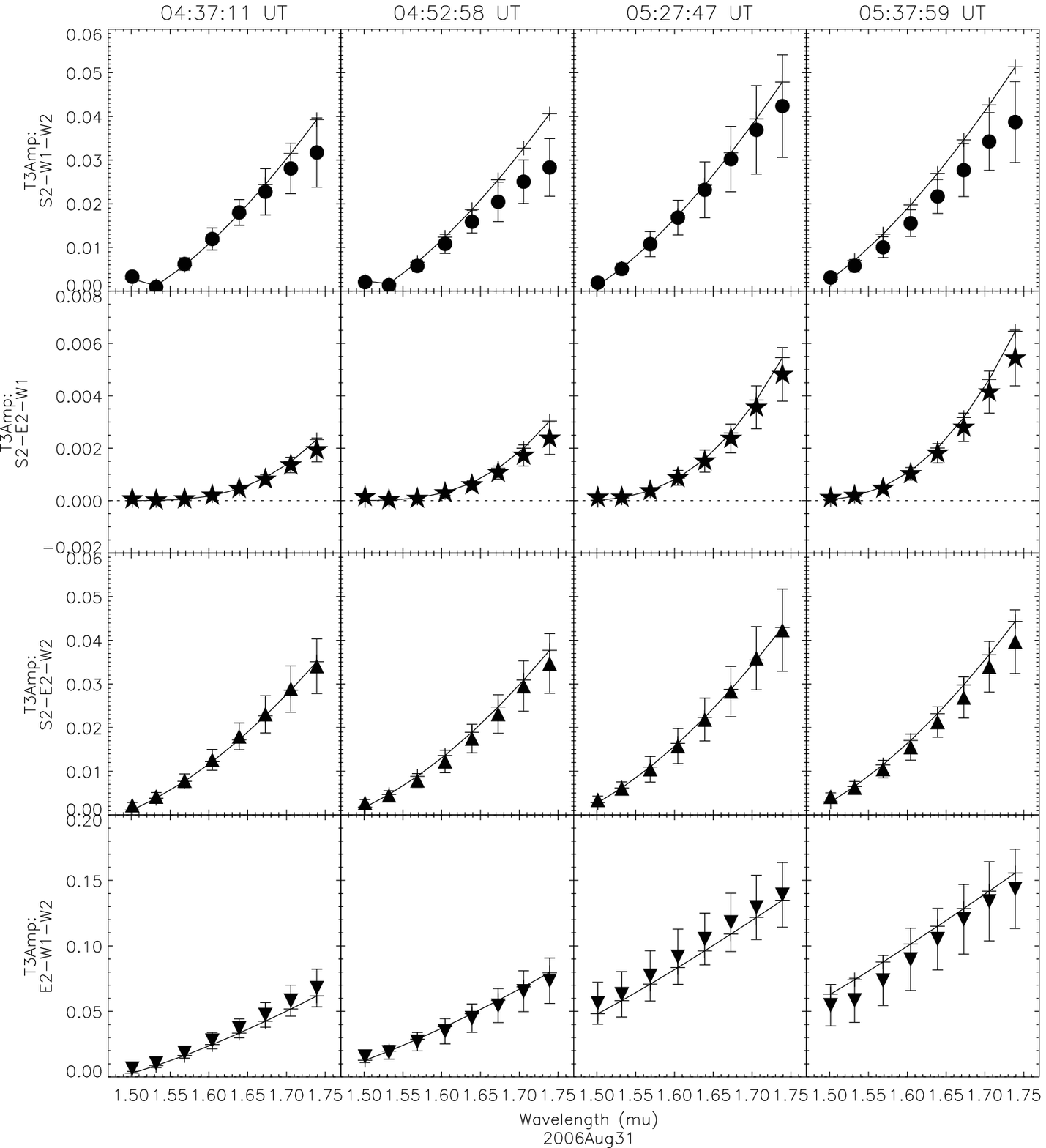}
\includegraphics[angle=0,width=3.2in, height= 3.8in]{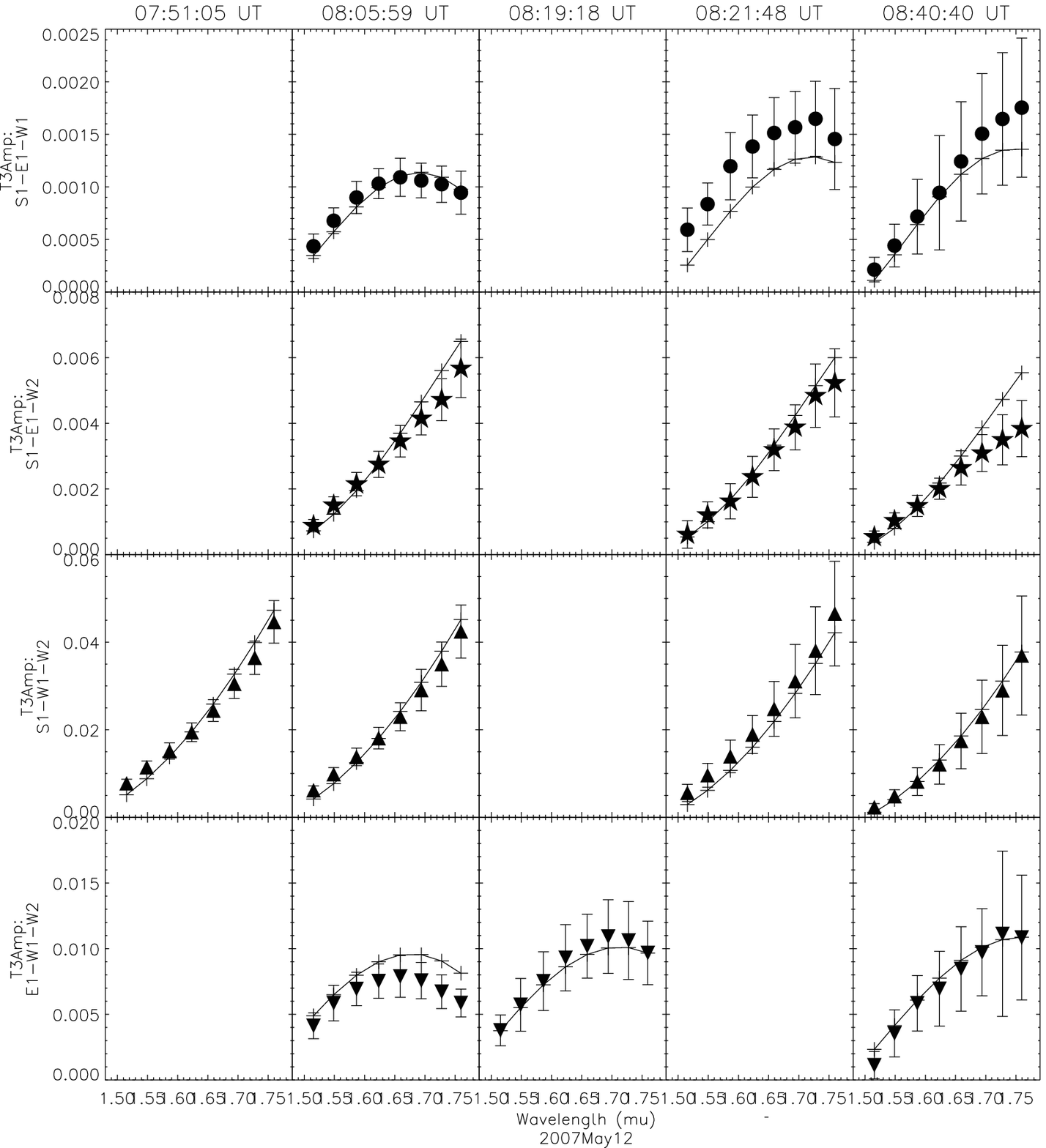}
}
\hphantom{.....}
\caption{ 
Similar to  Fig.\ref{alfoph_vis2} but showing the triple amplitudes for $\alpha$ Oph. The total $ \chi^2 _{\nu} $ for triple amplitude is 0.81.
 \label{alfoph_t3amp}}
\end{center}
\end{figure}

We also start with the standard gravity darkening model ($\beta$=0.25) for $\alpha$ Oph. We assume mass $= 2.10 \msun$ (see \S\ref{Teff}) and distance = 14.68 pc \citep{Gatewood2005} in the model. The metallicity $[Fe/H]$ of \alfoph is -0.16 \citep{Erspamer2003}, thus a Kurucz grid with metallicity of -0.2 is applied.
 Figure \ref{alfoph_model} shows the best-fit standard model of $\alpha$ Oph. The best-fit parameters are listed in Table \ref{alfoph_tab}. The associated errors of the parameters are also obtained using the bootstrap procedure  described in \S\ref{sec-alfcep}. 
The squared-visibilities, closure phases and triple amplitudes of the model are compared with  the data in Figures  \ref{alfoph_vis2}, \ref{alfoph_cp} and \ref{alfoph_t3amp}, respectively. The model shows that the photosphere of \alfoph is also elongated and has two bright polar areas and a dark equator. Its radius at the equator is $\sim20\%$ larger than at the poles. It is seen nearly equator-on with an inclination of $87\fdg70\pm0\fdg43$.
 The model also shows that \alfoph is rotating at 88.5\% of its break-up speed and the poles are $\sim1840$K hotter than the equator.

In the standard model, the ${\chi_{\nu}}^2$ of the closure phase only reaches 1.33 (Table \ref{alfoph_tab}), suggesting that we may need extra degrees of freedom to improve the fit. Therefore, following our approach for $\alpha$ Cep, we extend the standard model of \alfoph to a free $\beta$.
However,  although we have searched the parameter space extensively, we cannot find a unique $\beta$-free model for $\alpha$ Oph due to the same reason that  we encountered in imaging.
As we mentioned in \S\ref{imaging}, this issue stems from the near equator-on and symmetric brightness distribution of $\alpha$ Oph, causing the closure phases to be mostly $0^o$ or $\pm180^o$ (as shown in Fig.\ref{alfoph_cp}) and hence lack of enough non-zero signatures to constrain the model when $\beta$ is free. 

\begin{figure}[thb]
\begin{center}
{\includegraphics[angle=90,width=4in]{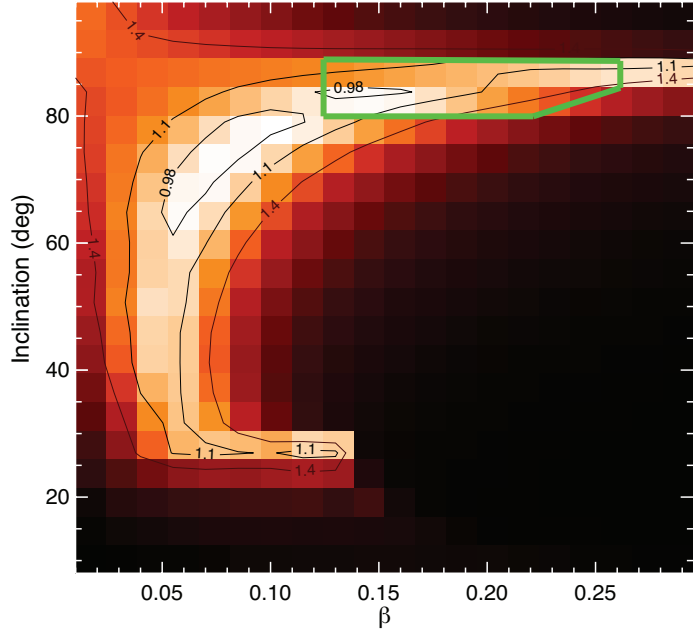}
}
\hphantom{.....}
\caption{ 
The  $\chi_{\nu}^2$ surface of $\beta$ and inclination  for $\alpha$ Oph. 
The corresponding probability is high throughout a large range of inclination and $\beta$, suggesting high degeneracy between the two parameters. 
The map also indicates the inclination at $\beta =0.25$ (i.e., the standard model) is well constrained and is nearly equator-on. Since the probability is dominated by the degeneracy effects of $\beta$ and inclination, we overplot the $\chi_{\nu}^2$ contours on the map instead of confidence intervals. The region enclosed in the green box has V$sin$i values inside the observed range of 210-240 \kms. The rest of the areas in the map fall outside the observed  V$sin$i range and thus can be ruled out, even though they may fit the data better. 
 \label{alpoph_ib}}
\end{center}
\end{figure}

Figure \ref{alpoph_ib} shows the $\chi_{\nu}^2$ space of inclination and $\beta$ for $\alpha$ Oph. Unlike the single peak of $\alpha$ Cep, \alfoph has several peaks spreading over a large range of inclination and $\beta$, indicating the inclination and  $\beta$ are highly degenerate and suggesting it is difficult to constrain a unique $\beta$-free model. 
Nevertheless,  the corresponding $V$sin$i$ values around the largest peak at $\beta\sim0.08$ fall outside the observed range of 210 - 240 \kms  (enclosed by the green box in Figure \ref{alpoph_ib}), suggesting the peak is not real but only due to the degeneracy of $\beta$ and inclination.
In addition, the peak around $\beta \sim 0.08$ corresponds to a fully convective star according to \citet{Lucy1967}. But it is unlikely for an A5 star to be fully convective, especially when its polar temperature is as high as 9300K. Therefore, we can rule out the largest peak around $\beta\sim0.08$. 
 Furthermore, the gravity darkening evolution models of \citet{Claret2000} show that the value of $\beta$ should be much larger than 0.15 for a $\sim 2\msun$ star with average T$_{eff}$ higher than 7500K, like $\alpha$ Oph.  The second peak around $\beta\sim0.15$ in Fig.\ref{alpoph_ib}, however, is not consistent with the models of \citet{Claret2000} although is inside the V$sin$i range. Thus, in this study we still prefer the other peak around the standard $\beta = 0.25$ model for $\alpha$ Oph.  
To break down the degeneracy and constrain the value of $\beta$ more accurately, we will need more observations with higher resolution, especially in the visible where limb-darkening and gravity darkening are more prominent.
  
\section{Physical properties and comparison with stellar evolution tracks}
\label{Teff}

In addition to the model parameters, we also calculate the true and apparent effective temperatures and luminosities for the two stars in Table \ref{alfcep_tab} \& \ref{alfoph_tab}. The true luminosity is estimated by integrating local $\sigma T_{eff} (\theta)^4$ (where $\sigma$ is the Stefan-Boltsman constant) over the stellar surface, and the true T$_{eff}$ is estimated from the total luminosity and the total surface area of the star. 
The apparent luminosity is obtained from $L=4\pi d^2 F_{bol}$, where the bolometric flux $F_{bol}$ is calculated by integrating the specific intensity over the whole spectrum and the projected angular area of the star. The apparent temperature is obtained from $ \sigma T_{eff}^4= \pi d^2 F_{bol} / A_{proj}$, where $A_{proj}$ is the projected area.

The  true T$_{eff}$  and luminosity of $\alpha$ Cep are very close to its apparent values due to its medium inclination (see Table \ref{alfcep_tab}).
Its true T$_{eff}$ from the $\beta$-free model is 7510 $\pm$ 160K, close to although slightly cooler than the $\sim$7700K estimate of VB06 and \citet{Gray2003}, as well as the 7740K estimate of \citet{Malagnini1990}. Its true luminosity is 18.1 $\pm$ 1.8 $\lsun$, consistent with the 17 $\lsun$ estimate from \citet{Malagnini1990} and the 17.3 $\lsun$ estimate of \citet{Simon1997}

The deviation of $\alpha$ Oph's true T$_{eff}$ and luminosity from its apparent values is very significant because of its near equator-on inclination.
Its true T$_{eff}$ from the standard model is estimated to be 8250$\pm$100K. Its apparent T$_{eff}$, on the other hand, is 7950K based on the model,  consistent with the apparent value of 7883 $\pm$ 63 K calculated by \citet{Blackwell1998} and  the value of $8030 \pm 160$ K by \citet{Malagnini1990}. Its apparent luminosity is 24.3 $\lsun$, in agreement with the 25.1 $\lsun$ value of \citet{Malagnini1990} but smaller than its true luminosity of 30.2 $\pm$ 1.3 $\lsun$.

\begin{figure}[thb]
\begin{center}
{\includegraphics[angle=90,width=3.2in]{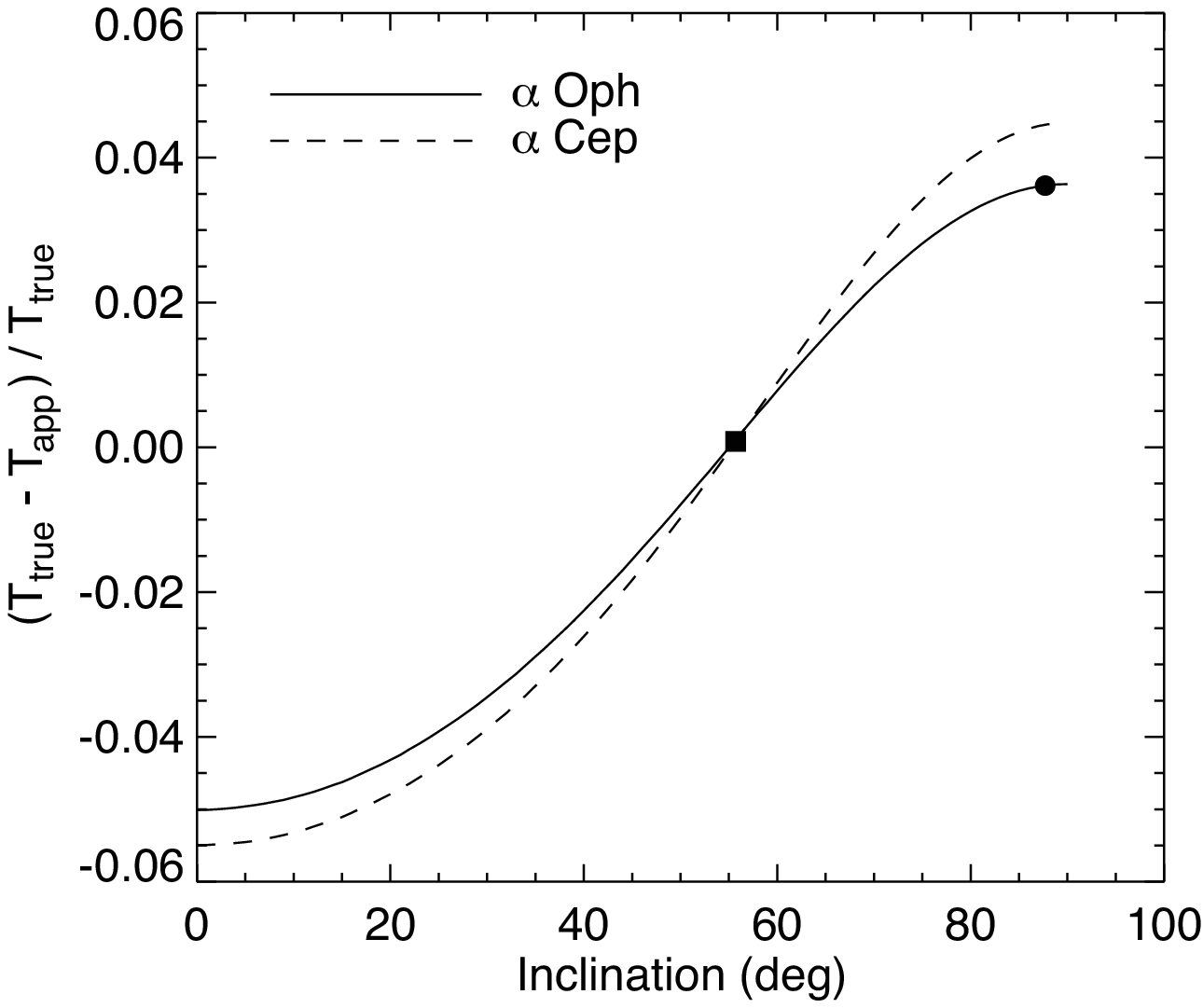}
\includegraphics[angle=90,width=3.2in]{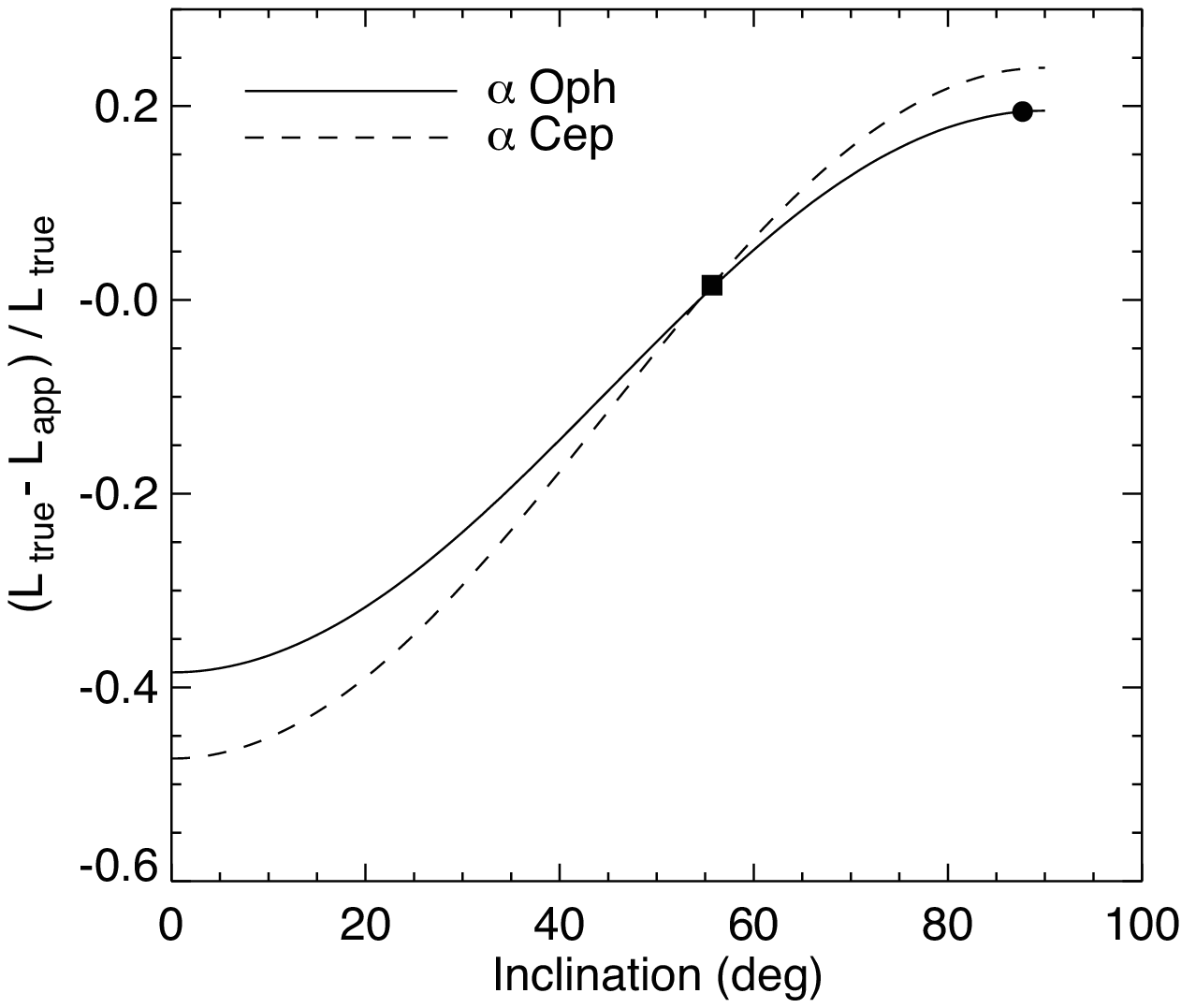}}
\hphantom{.....}
\caption{ 
Deviation of the apparent T$_{eff}$ and luminosity from  their true values at various inclinations. The solid line indicates the standard model ($\beta=0.25$) of $\alpha$ Oph. The dashed line indicates the $\beta=0.216$ model of $\alpha$ Cep. The apparent T$_{eff}$ and luminosity equal their true values at inclination of $\sim54^o$. \alfcep (filled square) is seen very close to this zero-difference value, but \alfoph (filled dot) is almost at the high end due to its large inclination.  
 \label{i_Teff}}
\end{center}
\end{figure}

Because rapid rotators are hotter at the poles and cooler at the equators, their apparent temperatures are therefore dependent on their inclinations, which can easily introduce large biases to the observed values. To investigate this effect, we plot in Figure \ref{i_Teff} the differences between the true and apparent values of T$_{eff}$s and luminosities as a function of inclination, scaled with their true values.  The plots show that when a star is inclined by $\sim54^o$, its apparent T$_{eff}$ and luminosity seen by the observers will be equal to their true values, just as  the case of $\alpha$ Cep and similar to the result of \citet{Gillich2008}. When the star is seen pole on, such as Vega \citep{Aufdenberg2006, Peterson2006b}, its apparent temperature can exceed the true value by $\sim5\%$, and the luminosity can exceed by $\sim40-50\%$ or even larger depending on the speed of the rotation, which explains the reason that Vega's luminosity was largely overestimated for a long time until recent studies of \citet{Aufdenberg2006} and \citet{Peterson2006b}. On the other hand, when a rapid rotator is equator-on, as the case of $\alpha$ Oph, its apparent temperature and luminosity can be underestimated by $\sim4\%$ and $\sim20\%$ respectively. The rotation speed of the star also affects the differences between its true and apparent values - the faster the star rotates, the larger the difference we see.

\begin{figure}[thb]
\begin{center}
{\includegraphics[angle=90,width=4in]{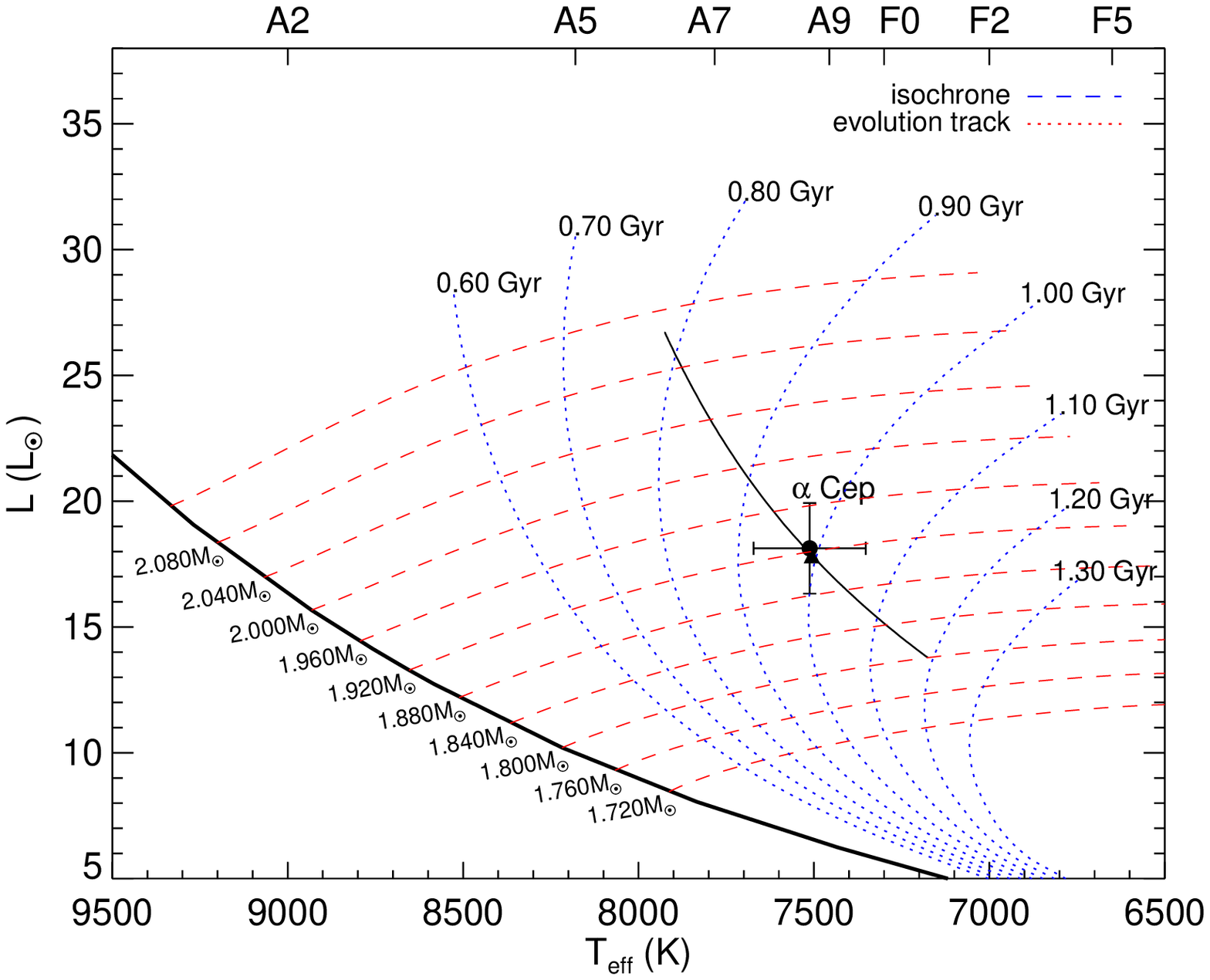}
\includegraphics[angle=90,width=4in]{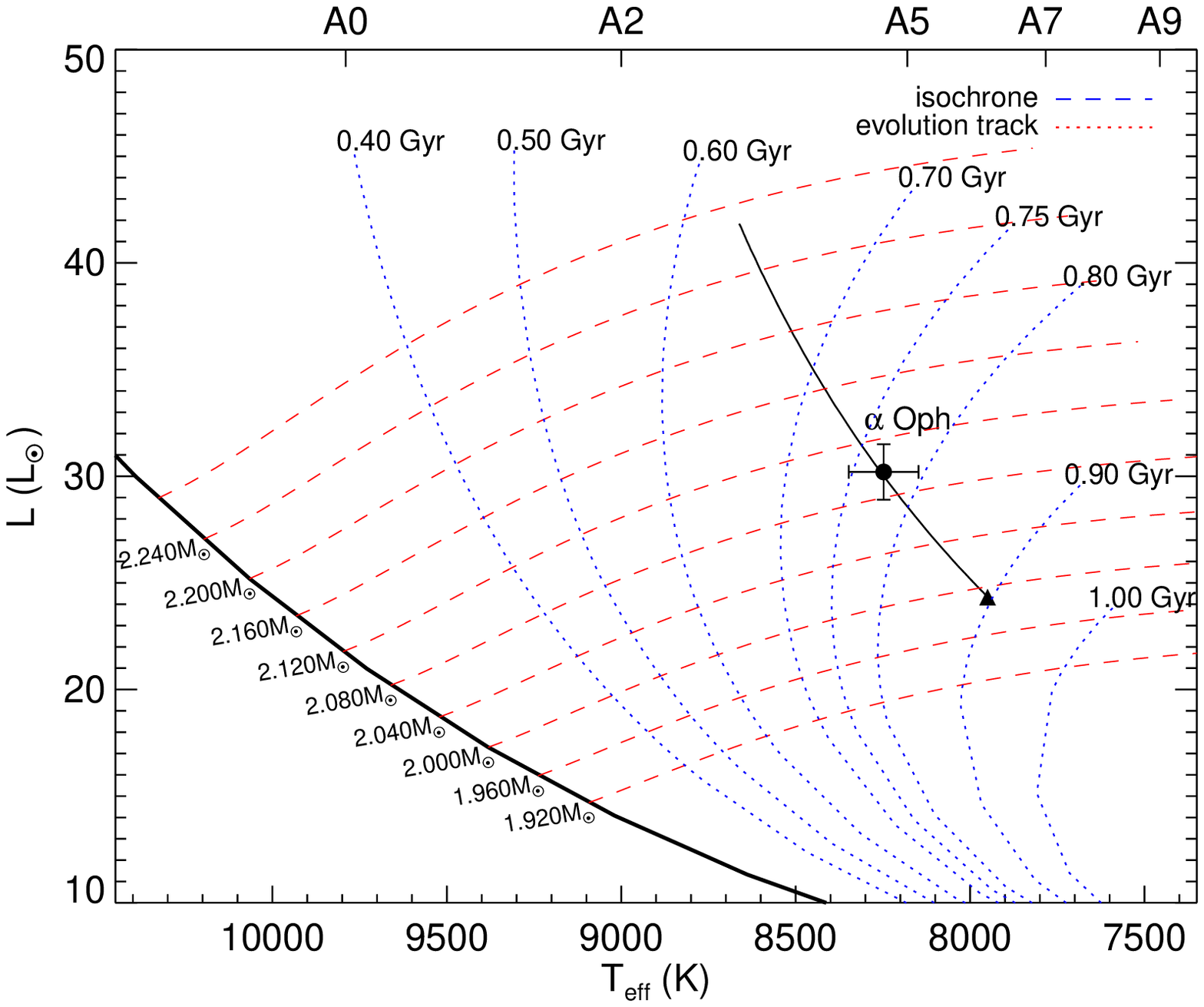}
}
\hphantom{.....}
\caption{  \scriptsize
Positions of \alfcep and \alfoph on the HR diagram. The top panel shows \alfcep and the $Y^{2}$ stellar models with Z$\sim0.02$. The bottom panel shows \alfoph and $Y^2$ stellar models with $Z\sim0.014$. 
The dashed red lines indicate the evolutionary tracks and the dotted blue lines indicate the isochrones. The filled dots with error bars indicate the true T$_{eff}$ and luminosity of the two stars, while the filled triangles indicate their apparent T$_{eff}$ and luminosity which are dependent on their inclinations. The solid lines that go through the points show the positions of \alfcep and \alfoph on the H-R diagram as a function of inclination \citep[also called ``inclination curves", ][]{Gillich2008}. These curves are more or less parallel to the Zero-Age Main Sequence indicated by the thick solid line at the bottom left of each plot, consistent with those of \citet{Gillich2008}. For a $90^o$ inclination, the positions of the stars will be at the lower end of the curve; and for a $0^o$ inclination, the stars will be at the higher end of the curves. These plots suggest the inclination of a star can significantly change its apparent location on the H-R diagram. 
 \label{hr}}
\end{center}
\end{figure}

Our estimates of the true T$_{eff}$s and luminosities of \alfcep and \alfoph also allow us to understand their current evolutionary status better. In Figure \ref{hr} we plot the H-R diagram and the corresponding $Y^2$ stellar evolution tracks and isochrones \citep{Demarque2004} for \alfcep and $\alpha$ Oph. Their possible ranges of locations on the H-R diagram \citep[also called ``inclination curve",][]{Gillich2008} are also shown in the plots.
The top panel shows that \alfcep appears to be an A9 type star on the H-R diagram based on its apparent temperature and luminosity (filled triangle). However, it is classified as an A8V star by \citet{Gray2003}, earlier than that inferred from the top panel.
Similarly, in the bottom panel of Figure \ref{hr}, \alfoph  appears roughly as an A6.5 type star. Its apparent spectral type from \citet{Gray2001} is A5IV, also earlier than that inferred from the figure.  We infer that this is because the spectra of the two stars are dominated by spectral lines from  the hotter and brighter polar regions, causing their overall spectral classification to be biased toward the types of their poles which appear earlier than other regions of the stars. Therefore, for the case of an equator-on star, such as $\alpha$ Oph, although its apparent effective temperature is lower than its true temperature due to the inclination, its spectral type derived from spectroscopy can compensate this effect and make it look closer to its true spectral type. However, for a pole-on star such as Vega, this bias can not  be compensated, and the spectral types derived from both spectroscopy and apparent temperature will appear earlier than its true type. 
This phenomenon indicates that the spectral types of rapid rotators are not only biased by their inclinations, but also by  the spectral lines of their polar regions.

Using the $Y^2$ models, we estimate that  \alfcep  has a mass of $1.92\pm0.04 \msun$, slightly smaller than the estimate of VB06. Its age is estimated to be $0.99\pm0.07$ Gyrs. 
We also estimate that \alfoph has a mass of $2.10\pm 0.02\msun$, and an age of $0.77\pm0.03$ Gyrs. Its apparent position in the H-R diagram, however, indicates a lower mass of 1.99$\msun$, which is again consistent with the 2.0$\msun$ estimate of \citet{Malagnini1990} and \citet{Augensen1992}. However, this value is much lower than the 2.84$\msun$ value of \citet{Gatewood2005} and the 4.9\msun~ of \citet{Kamper1989}.  To address the differences, we derive the mass range of \alfoph using our new method of estimating mass in the next section (\S\ref{new_method}), and conclude the result of \citet{Gatewood2005} and \citet{Kamper1989} can be ruled out. The estimated masses and ages of \alfcep and \alfoph are included in Tables \ref{alfcep_tab} and \ref{alfoph_tab} respectively.

We note that the $Y^2$ models are for non-rotating stars, whereas both \alfcep and \alfoph are rapid rotators. The fact that rotation may extend the  main-sequence lifetime \citep{Kiziloglu1996, Maeder2000} implies that our age estimates may not be accurate and needs further investigation.
We also note that the masses of \alfcep and \alfoph are both estimated based on non-$\alpha$-enhanced $Y^2$ models. Studies have shown that rapid rotation can change the abundance of a star \citep[e.g.,][]{Pinsonneault1997} and enhance the $\alpha$-rich elements \citep{Yoon2008}, resulting in very different estimates of its mass and age. Hence to derive the masses of \alfcep and \alfoph more accurately, detailed abundance studies are required to determine if they are $\alpha$-enhanced and what abundance to use for their evolutionary models.

\section{A new method to estimate the mass of a star}
\label{new_method}
Mass is the most fundamental property of stars. The determination of stellar masses mostly relies on orbital measurements of binary systems \citep[e.g.,][]{Zhao2007}, stellar evolution models together with measurements of other stellar properties \citep[e.g.,][]{vanBelle2006}, and asteroseismology together with measurements of stellar radii \citep[e.g.,][]{Creevey2007}.  Here we propose a new method to estimate the mass of a star based on our modeling of rapid rotators.

Since we can determine the inclination, equatorial radius and the fractional rotation speed of a rapid rotator from our model, we therefore can combine the model of a rapid rotator with its mass to estimate the equatorial velocity and the V$\sin i$ value.  We can also reverse the process,  taking a precise measurement of V$\sin i$ and a best-fit rotator model to determine the mass of a star. This approach is most suitable for radiative rapid rotators which can be interpreted by the standard gravity darkening model, and also non-fully-radiative rotators if a more sophisticated fluid model is constructed \citep[e.g.,][]{Jackson2004, MacGregor2007, Espinosa2007}. For stars with less accurate models, we can also use this method to roughly estimate their masses. The precision of V$\sin i$ is also crucial for a precise mass estimate. As a preliminary test, we first apply this method to \alfcep and $\alpha$ Oph. 

The $Vsin$i range of \alfcep (180 \kms - 245 \kms) corresponds to a large mass range of 1.3$\msun$ to 2.4$\msun$~ based on the $\beta$-free model in \S\ref{sec-alfcep}. The mass of \alfcep determined from stellar models, on the other hand, is $1.92\pm0.04 \msun$~(see \S\ref{Teff}), well within the mass range given by $Vsin$i.
Similarly, the $Vsin$i range of \alfoph (210 \kms - 240 \kms) gives a mass range of 1.7\msun~ to 2.2\msun~ when combined with the model in \S\ref{sec-alfoph}.  
Its mass determined from stellar models, $2.1\pm 0.02\msun$~ (see \S\ref{Teff}), is also within the range.
 By contrast, the study of \citet{Gatewood2005} and \citet{Kamper1989} gave a mass of 2.84\msun~ and 4.9\msun~ to $\alpha$ Oph respectively, far outside the range given by $Vsin$i, and hence can be ruled out. Since \alfoph is also a known astrometric binary, it is the ideal target to further test this new method by comparing its mass with that determined from the astrometric orbit. We are currently pursuing this study (Oppenheimer et al.2008, private communication) and will also present it in a future work.

\section{Discussion}
\label{discuss}

Although the $\beta$-free model of \alfcep is consistent with the synthesized image (Fig.\ref{alfcep_img}) in basic features such as the bright pole and the dark equator, we also notice that the equator of the image is darker and cooler than that of the model - a phenomenon seen in a previous study of Altair \citep{Monnier2007sci}. The existence of the darker-than-expected equator on both stars implies that  the extra gravity darkening may be real. However, it can also be due to a systematic effect of the imaging program. To confirm this conclusion we will need further studies such as model-independent latitudinal temperature profiles.

Our models show that both \alfcep and \alfoph have polar temperatures well above 8000K and equatorial temperatures below 7500K, which means, according to the stellar atmospheric grid of \citet{Kurucz1979}, the polar areas of \alfcep and \alfoph are radiative and their equators can have convections, especially for \alfcep as its equatorial temperature is lower than that of $\alpha$ Oph. Since the existence of convection tends to lower the value of the average gravity darkening coefficient $\beta$ of the whole star \citep{Claret1998}, it may be the cause of $\beta<0.25$  in the $\beta$-free model of $\alpha$ Cep. The unusually strong chromosphere activity of \alfcep among A stars \citep{Walter1995, Simon1997} also provides evidence to the convective layers since the chromosphere is directly linked to magneto-convection.  Another A star with strong chromosphere activities, Altair, is also a rapid rotator spinning at  92\% of its break-up speed and has an equatorial temperature of 6860K \citep{Monnier2007sci}. This suggests that although A stars are generally considered to have no chromospheres due to their very thin or lack of convective layers \citep{Simon2002}, rapid rotators may have exceptions at their equators due to gravity darkening. This is also consistent with the conclusion from the hydrodynamic model of \citet{Espinosa2007}. This effect may also shed some light on the searches for the onset of chromosphere and the transition from radiative to convective envelopes among early type stars \citep[e.g.,][]{Simon2002}.

Since convection also tends to smear out the temperature differences between the hot and cool regions of the stellar surface and make their intensity contrast lower, other mechanisms such as differential rotation \citep[e.g.,][]{Espinosa2007} may also exist in the equators of these stars in order to make the equator darker and cooler as in the image. For instance, a faster differentially spinning equator  will have stronger gravity darkening, thus will appear darker than that of the standard model. However, the darker equator, if it is real, can also be caused by a very different form of gravity darkening law. To further address this issue, we will need detailed line profile studies and images at visible since gravity darkening is more prominent in the visible than in the $H$ band.

 The $87\fdg70$ inclination of \alfoph differs from its orbital inclination  by about $27^o$
 \citep[$i\sim115^o$,][]{Kamper1989, Augensen1992, Gatewood2005}, 
 indicating the spin of \alfoph is not coplanar with its orbit. 
 Even more interesting, the orbit of the binary is highly eccentric ($e\sim0.8$, Kamper et al. 1989 and Gatewood 2005; $e=0.57$, Augensen \& Heintz 1992), implying the non-coplanarity and the high eccentricity of the system may be related to each other through interactions of the two stars with their disks in their early formation stages.


\section{Conclusion}
\label{conclusion}
We have modeled the surface brightness distributions of \alfcep and \alfoph using the gravity darkening model. We have also reconstructed an aperture synthesis image for $\alpha$ Cep, but no reliable image for \alfoph is available due to its lack of closure phase signatures caused by its nearly symmetric brightness distribution.
The image of \alfcep shows the star is oblate and its equator is darker than its poles, directly confirming the gravity-darkening phenomenon. The models show that both stars are rotating close  to their break-up speed. They both appear oblate and have large latitudinal temperature gradient due to gravity darkening. 
A standard gravity darkening model of $\beta$=0.25 is adopted for $\alpha$ Oph, and its inclination is determined to be $87.70^o$.  For $\alpha$ Cep, a $\beta=0.216$ model fits the data better and also agrees better with the image. It has a medium inclination angle of $55.70^o$.

Our models also allow us to calculate and compare the true T$_{eff}$s and luminosities of the two stars with their apparent values. 
We show that \alfoph has a true T$_{eff}$ of 8250K and luminosity of 30.2 $\lsun$, significantly larger than its apparent values due to its equator-on inclination. 
The true T$_{eff}$ and luminosity of $\alpha$ Cep, on the other hand, appear very close to its apparent values because of its medium inclination. 
The spectral classification of the two stars from literatures, however, suggests earlier spectral types for both stars than that derived from their apparent T$_{eff}$s and luminosities. We infer that this is because the spectra of the two stars are dominated by lines from their hotter and brighter polar regions which appear much earlier in spectral type than the other regions of the stars, causing their overall spectral classification to be biased toward their polar areas. 

The temperatures and luminosities in turn allow us to make rough estimates of the masses of the two stars through stellar evolution models. The mass of \alfcep is estimated to be  $1.92\msun$, and the mass of \alfoph is $2.10\msun$. However, due to possible abundance anomaly caused by rapid rotation, the exact masses of the two stars still have to be scrutinized when a detailed abundance analysis is available. 

Our gravity darkening models also allow us to propose a new method to estimate the masses of rapid rotators together with precise measurements of V$sin$i. We have tested this method on both stars and found our mass estimate from the stellar models are within the range. The star \alfoph will be a good target  to further test this method as it is also an astrometric binary.

Our models show that the equatorial temperatures of  \alfoph and especially \alfcep are low enough to meet the onset conditions of convection, implying that convections in the equatorial region can be a reason of the unusually high chromosphere activities of $\alpha$ Cep. 
Although the \alfcep model agrees with its image in general, the image shows extra darkening at the equator which is not expected by our gravity darkening model but is consistent with the previous result of Altair. This effect, if is real, is most likely caused by differential rotation of the star. But to further confirm the conclusion, detailed high resolution line profile analysis and images at visible are needed.

\acknowledgments

We thank Michael Ireland for the MACIM package used in
this work. We thank the valuable discussions with C. Cowley.
We also thank the referee for valuable suggestions and comments. 
The CHARA Array is funded by the National
Science Foundation through NSF grants AST-0307562 and
AST-0606958 and by the Georgia State University. We thank
the support for this work by the Michelson Graduate Fellowship
(M. Z.), the NSF grants NSF-AST
0352723, NSF-AST 0707927, NASA NNG 04GI33G (J. D. M.),
and EU grant MOIF-CT-2004-002990 (N. T.). E. P. was formally
supported by the Michelson Postdoctoral Fellowship and
is currently supported by a Scottish Universities Physics Association
(SUPA) advanced fellowship. This work has made use of the PTI data archive maintained by the Michelson Science Center.
\clearpage

\begin{deluxetable}{llclc}
\tabletypesize{\scriptsize}
\tablecaption{Observation logs for \alfoph~ and \alfcep}
\tablewidth{0pt}
\tablehead{
\colhead{Target} & \colhead{Obs. Date} & \colhead{Telescopes} & \colhead{Calibrators} & \colhead{Chopper} \\}
\startdata
\alfoph & UT 2006Jun20   & W1-W2-S1-S2  & $\alpha$ Sge & no\\
             & UT 2006Jun21	& W1-W2-S1-S2  & $\zeta$ Oph, $\gamma$ Ser & no \\ 
             & UT 2006Aug28	& S2-E2-W1-W2  & $\upsilon$ Peg & no \\
             & UT 2006Aug29	& S2-E2-W1-W2  & $\gamma$ Lyr, $\upsilon$ Peg & no \\ 
             & UT 2006Aug30	& S2-E2-W1-W2  & $\gamma$ Lyr &yes \\
             & UT 2006Aug31	& S2-E2-W1-W2  & $\gamma$ Lyr, $\upsilon$ Peg & yes \\
             & UT 2007May10  & S1-E1-W1-W2  & $\zeta$ Oph, $\tau$ Aql & yes \\
             & UT 2007May12  & S1-E1-W1-W2  & $\zeta$ Oph, $\tau$ Aql & yes \\
\hline
\hline
\\
\alfcep	&UT 2006Oct09 &	S2-E2-W1-W2	& 29 Peg, $\upsilon$ And, $\zeta$ Per & yes \\
		&UT 2006Oct11 &	S2-E2-W1-W2	& $\upsilon$ And, $\zeta$ Per & yes \\
		&UT 2006Oct12 &	S2-E2-W1-W2	& 29 Peg,  $\zeta$ Per & yes \\
		&UT 2006Oct16 &	S2-E2-W1-W2	& 29 Peg, $\upsilon$ And & yes \\
\enddata
\label{obslog}
\end{deluxetable}

\begin{deluxetable}{lcl}
\tabletypesize{\scriptsize}
\tablecaption{Calibrator diameters}
\tablewidth{0pt}
\tablehead{
\colhead{Calibrator} & \colhead{UD diameter ($mas$)} & \colhead{Reference} 
}
\startdata
$\alpha$ Sge & 1.32 $\pm$0.02 & Uniform-disk fit to PTI archive data\tablenotemark{a} \\
$\zeta$ Oph   & 0.51 $\pm$ 0.05 & \cite{Hanbury1974} \\
$\gamma$ Ser & 1.21$\pm$ 0.05& Uniform-disk fit to PTI archive data  \\
$\gamma$ Lyr &  0.74 $\pm$0.10& \cite{Leggett1986} \\
$\upsilon$ Peg & 1.01 $\pm$ 0.04 & \cite{Blackwell1994}\\
$\tau$ Aql & 1.10 $\pm$ 0.01& \citet{Merand2005, Merand2006}\\
29 Peg&1.0 $\pm$ 0.1 & MIRC measurement \\
$\upsilon$ And & 1.17 $\pm$ 0.02 & Boden 2008\tablenotemark{b} \\
$\zeta$ Per &0.67 $\pm$ 0.03 & getCal\tablenotemark{c} \\
\enddata
\tablenotetext{a}{available at http://mscweb.ipac.caltech.edu/mscdat-pti}
\tablenotetext{b}{SED fit, private communication}
\tablenotetext{c}{http://mscweb.ipac.caltech.edu/gcWeb/gcWeb.jsp}

\label{cals}
\end{deluxetable}

\begin{deluxetable}{lcc}
\tabletypesize{\scriptsize}
\tablecaption{Best-fit and physical parameters of \alfcep}
\tablewidth{0pt}
\tablehead{ 
\colhead{Model Parameters} & \colhead{Standard ($\beta=0.25$) } 
&\colhead{Non-standard ($\beta$-free)\tablenotemark{*}} }
\startdata
Inclination (degs)         &   64.91 $\pm$ 4.11         & 55.70  $\pm$ 6.23\\
Position Angle (degs) & -178.26$\pm$ 4.10       & -178.84 $\pm$ 4.28\\
T$_{pol}$ (K)                &    8863$\pm$ 260             & 8588 $\pm$ 300\\
R$_{pol}$ ($\rsun$)    &   2.199 $\pm$ 0.035          & 2.162 $\pm$ 0.036\\
T$_{eq}$  (K)                &   6707 $\pm$ 200              &6574 $\pm$ 200\\
R$_{eq}$  ($\rsun$)    &    2.739$\pm$ 0.040       & 2.740 $\pm$ 0.044\\
$\omega$                     &   0.926 $\pm$ 0.018         & 0.941 $\pm$0.020 \\
$\beta$                          &   0.25 (fixed)                      & 0.216$\pm$ 0.021\\
Model V Magnitude\tablenotemark{a}     &   2.45                                 &2.45\\
Model H Magnitude\tablenotemark{b}     &   1.92                                &1.91 \\
Model v $\sin i$ (km/s)  &        237                                                             & 225\\
\hline
Total $\chi^2_{\nu}$       & 1.21                                      &1.18       \\
Vis$^2$ $\chi^2_{\nu}$ & 0.79                                   & 0.80 \\
CP $\chi^2_{\nu}$          & 1.43                                     & 1.27 \\
T3amp $\chi^2_{\nu}$   & 1.71                                     &1.76 \\
\hline
\hline
\colhead{Other Physical Parameters}\\
\hline
True T$_{eff} (K)$ & 7690 $\pm$ 150 & 7510 $\pm$ 160\\
True Luminosity (\lsun) & 20.1 $\pm$ 1.6 & 18.1 $\pm$ 1.8\\
Apparent T$_{eff} (K)$ & -  & 7510 \\
Apparent Luminosity (\lsun) & - & 17.9 \\
Mass (\msun)\tablenotemark{c} &  - & 1.92 $\pm$ 0.04 \\
Age (Gyrs)\tablenotemark{c} &-  &      0.99 $\pm$ 0.07  \\
$[Fe/H]$\tablenotemark{d}  & \multicolumn{2}{c}{0.09} \\
Distance (pc)\tablenotemark{e} & \multicolumn{2}{c}{14.96}\\
\enddata
\tablenotetext{*}{The $\beta$-free model is adopted as the final model, see text of \S\ref{sec-alfcep} for detail.}
\tablenotetext{a}{V magnitude from literature: 2.456 $\pm$ 0.002 \citep{Perryman1997}  }
\tablenotetext{b}{H magnitude from literature: 2.13 $\pm$ 0.18 \citep{Cutri2003}}
\tablenotetext{c}{Based on the $Y^2$ stellar evolution model \citep{Demarque2004}.}
\tablenotetext{d}{\cite{Gray2003} }
\tablenotetext{e}{\cite{Perryman1997} }
\label{alfcep_tab}
\end{deluxetable}

\begin{deluxetable}{lc}
\tabletypesize{\scriptsize}
\tablecaption{Best-fit and physical parameters of \alfoph}
\tablewidth{0pt}
\tablehead{ 
\colhead{Model Parameters} & \colhead{Standard  ($\beta=0.25$)} }
\startdata
Inclination (degs)          &  87.70  $\pm$ 0.43             \\
Position Angle (degs)  &  -53.88 $\pm$ 1.23             \\
T$_{pol}$ (K)                 &  9300   $\pm$ 150              \\
R$_{pol}$ ($\rsun$)     &  2.390  $\pm$ 0.014           \\
T$_{eq}$  (K)                 &  7460   $\pm$ 100              \\
R$_{eq}$  ($\rsun$)    &  2.871  $\pm$ 0.020            \\
$\omega$                     &  0.885  $\pm$ 0.011            \\
$\beta$                          &  0.25 (fixed)                          \\
Model V Magnitude\tablenotemark{a}     & 2.086     \\
Model H Magnitude\tablenotemark{b}     & 1.66       \\
Model v $\sin i$ (km/s)                                  & 237        \\
\hline
Total $\chi^2_{\nu}$                                   & 0.91       \\
CP $\chi^2_{\nu}$                                      & 1.33        \\
Vis$^2$ $\chi^2_{\nu}$                             & 0.72        \\
T3amp $\chi^2_{\nu}$                               & 0.81        \\
\hline
\hline
\colhead{Other Physical Parameters} \\
\hline
True T$_{eff} (K)$ & 8250 $\pm$ 100 \\  	  
True Luminosity (\lsun) & 30.2 $\pm$ 1.3 \\  
Apparent T$_{eff} (K)$ & 7950  \\ 
Apparent Luminosity (\lsun) & 24.3 \\
Mass (\msun)\tablenotemark{c} & 2.10 $\pm$ 0.02 \\
Age (Gyrs)\tablenotemark{c} & 0.77 $\pm$ 0.03  \\
$[Fe/H]$\tablenotemark{d}  & -0.16 \\
Distance (pc)\tablenotemark{e} & 14.68\\
\enddata
\tablenotetext{a}{V magnitude from literature: 2.086 $\pm$ 0.003  \citep{Perryman1997}}
\tablenotetext{b}{H magnitude from literature: 1.66  $\pm$ 0.03   \citep[weighted average of fluxes from: ][]{Alonso1998, Cohen1999,  Cutri2003}}
\tablenotetext{c}{Based on the $Y^2$ stellar evolution model \citep{Demarque2004}.}
\tablenotetext{d}{\cite{Erspamer2003} }
\tablenotetext{e}{\cite{Gatewood2005} }
\label{alfoph_tab}
\end{deluxetable}



\bibliographystyle{apj}  
\bibliography{alfoph_v3}  

\end{document}